\RequirePackage{lineno}
\documentclass[aps,prb,10pt,tightenlines,superscriptaddress,nofootinbib,longbibliography]{revtex4-2}

\usepackage{bbold} 
\usepackage{bm}
\usepackage{graphicx}
\usepackage{placeins} 
\usepackage{braket}
\usepackage{epstopdf}
\usepackage{color}
\usepackage{subcaption}
\usepackage[colorlinks,pdfencoding=auto,psdextra]{hyperref}
\usepackage[dvipsnames]{xcolor} 
\usepackage{amsmath} 
\usepackage{amssymb} 
\usepackage{ulem} 
\usepackage[T2A]{fontenc}
\usepackage[cp1251]{inputenc}
\usepackage[english]{babel}

\def\emph{\textit}

\def\be{\begin{equation}}
\def\ee{\end{equation}}
\def\bea{\begin{eqnarray}}
\def\eea{\end{eqnarray}}

\usepackage{empheq}
\definecolor{myblue}{rgb}{.8, .8, 1} 
    
\usepackage[bbgreekl]{mathbbol}

\renewcommand{\S}{\mathop{\mathcal S}}
\newcommand{\X}{\mathop{\mathcal X}}
\newcommand{\Y}{\mathop{\mathcal Y}}
\newcommand{\Z}{\mathop{\mathcal Z}}

\newcount\mycounter

\newcommand{\add}[1]{\textcolor{black}{#1}}

\begin{document}


\author{E. Kirstein}
\affiliation{Experimentelle Physik 2, Technische Universit\"{a}t Dortmund, 44227 Dortmund, Germany}

\author{D. R. Yakovlev}
\affiliation{Experimentelle Physik 2, Technische Universit\"{a}t Dortmund, 44227 Dortmund, Germany} 
\affiliation{Ioffe Institute, Russian Academy of Sciences, 194021 St. Petersburg, Russia}

\author{M. M. Glazov}
\affiliation{Ioffe Institute, Russian Academy of Sciences, 194021 St. Petersburg, Russia}

\author{E. A. Zhukov}
\affiliation{Experimentelle Physik 2, Technische Universit\"{a}t Dortmund, 44227 Dortmund, Germany}
\affiliation{Ioffe Institute, Russian Academy of Sciences, 194021 St. Petersburg, Russia}

\author{D. Kudlacik}
\affiliation{Experimentelle Physik 2, Technische Universit\"{a}t Dortmund, 44227 Dortmund, Germany}

\author{I. V. Kalitukha}
\affiliation{Ioffe Institute, Russian Academy of Sciences, 194021 St. Petersburg, Russia}

\author{V. F. Sapega}
\affiliation{Ioffe Institute, Russian Academy of Sciences, 194021 St. Petersburg, Russia}

\author{G. S. Dimitriev}
\affiliation{Ioffe Institute, Russian Academy of Sciences, 194021 St. Petersburg, Russia}

\author{M. A. Semina}
\affiliation{Ioffe Institute, Russian Academy of Sciences, 194021 St. Petersburg, Russia}

\author{M. O. Nestoklon}
\affiliation{Ioffe Institute, Russian Academy of Sciences, 194021 St. Petersburg, Russia}

\author{E. L. Ivchenko}
\affiliation{Ioffe Institute, Russian Academy of Sciences, 194021 St. Petersburg, Russia}

\author{N. E. Kopteva}
\affiliation{Experimentelle Physik 2, Technische Universit\"{a}t Dortmund, 44227 Dortmund, Germany}

\author{D. N. Dirin}
\affiliation{Department of Chemistry and Applied Biosciences,
Laboratory of Inorganic Chemistry, ETH Z\"{u}rich, 8093 Z\"{u}rich, Switzerland}

\author{O. Nazarenko}
\affiliation{Department of Chemistry and Applied Biosciences,
Laboratory of Inorganic Chemistry, ETH Z\"{u}rich, 8093 Z\"{u}rich, Switzerland}

\author{M. V. Kovalenko}
\affiliation{Department of Chemistry and Applied Biosciences,
Laboratory of Inorganic Chemistry, ETH Z\"{u}rich, 8093 Z\"{u}rich, Switzerland} 
\affiliation{Department of Advanced Materials and
Surfaces, Laboratory for Thin Films and Photovoltaics, Empa - Swiss Federal Laboratories for Materials Science and Technology, 8600 D\"{u}bendorf, Switzerland}

\author{A. Baumann}
\affiliation{Experimental Physics VI, Julius-Maximilian University of W\"urzburg, 
97074 W\"{u}rzburg, Germany}

\author{J. H\"ocker}
\affiliation{Experimental Physics VI, Julius-Maximilian University of W\"urzburg, 
97074 W\"{u}rzburg, Germany}

\author{V. Dyakonov}
\affiliation{Experimental Physics VI, Julius-Maximilian University of W\"urzburg, 
97074 W\"{u}rzburg, Germany}

\author{M. Bayer}
\affiliation{Experimentelle Physik 2, Technische Universit\"{a}t Dortmund, 44227 Dortmund, Germany} 
\affiliation{Ioffe Institute, Russian Academy of Sciences, 194021 St. Petersburg, Russia}

\title{The Land\'e factors of electrons and holes in lead halide perovskites: \\ universal dependence on the band gap}

\date{\today}

\begin{abstract}
The Land\'e or $g$-factors of charge carriers are decisive for the spin-dependent phenomena in solids and provide also information about the underlying electronic band structure. We present a comprehensive set of experimental data for values and anisotropies of the electron and hole Land\'e factors in hybrid organic-inorganic (MAPbI$_3$, \add{MAPb(Br$_{0.5}$Cl$_{0.5}$)$_3$, MAPb(Br$_{0.05}$Cl$_{0.95}$)$_3$, FAPbBr$_3$}, FA$_{0.9}$Cs$_{0.1}$PbI$_{2.8}$Br$_{0.2}$) and all-inorganic (CsPbBr$_3$) lead halide perovskites, determined by pump-probe Kerr rotation and spin-flip Raman scattering in magnetic fields up to 10~T at cryogenic temperatures. Further, we use first-principles DFT calculations in combination with tight-binding and $\bm k \cdot \bm p$ approaches to calculate microscopically the Land\'e factors. The results demonstrate their universal dependence on the band gap energy across the different perovskite material classes, which can be summarized in a universal semi-phenomenological expression, in good agreement with experiment. 
\end{abstract}

\maketitle


Lead halide perovskite materials have attracted huge attention in recent years due to their exceptional electronic and optical characteristics, which make them highly promising for various applications in photovoltaics~\cite{jeong2021,NRELchart}, optoelectronics~\cite{hansell_c2019,eperon_formamidinium_2014,wang_spin-optoelectronic_2019,fu_metal_2019,jena2019,piveteau2020}, X-ray detectors \cite{nazarenko2017,wei_halide_2019}, etc. Their chemical formula $A$Pb$X_3$ where the cation $A=$ cesium (Cs), methylammonium (MA), formamidinium (FA) and the anion $X=$ Cl, Br, I, offers a huge flexibility in composition making the band gap tunable from the infrared up to ultraviolet spectral range. Interestingly, the perovskite band structure is inverted compared to common III-V and II-VI semiconductors. As a result, the strong spin-orbit interaction influences mostly the conduction band rather than the valence band. Also strong Rashba spin splittings have been predicted both for the valence and conduction bands~\cite{PhysRevLett.117.126401}. 

Detailed studies of the band structure require a concerted effort of suitable experimental and theoretical approaches. The low charge carrier mobility~\cite{Herz:2017ul} hampers methods like electrical transport, also in magnetic field, and cyclotron resonance, which are usually applied to study the band structure of solids, while ion diffusion obstructs the application of capacity-based methods. Angle resolved photoemission spectroscopy~\cite{lee2017,zu2019,yang2021} provides promising results but, so far, with insufficient accuracy. Optics in strong magnetic fields gives access to carrier effective masses and exciton features~\cite{Miyata2015,Baranowski2020}, where the parameter values can be, however, influenced by the field through band mixing. 

Spin physics provides high precision tools for addressing the electronic states in the vicinity of the band gap. Namely, the Land\'e or $g$-factors of electrons and holes are inherently linked via their values and anisotropies to the band parameters, which also determine the charge carrier effective masses~\cite{ivchenko2005,yu2016}. On the other hand, the Land\'e factors are the key parameters for the coupling of spins to a magnetic field and thus govern related basic phenomena and spintronics applications, which belong to a largely uncharted area for perovskites. The first concise reports show great promise demonstrating optical orientation~\cite{giovanni2015,nestoklon2018,wang2018,wang_spin-optoelectronic_2019,pan2020} and optical alignment~\cite{nestoklon2018}, polarized emission in magnetic field~\cite{zhang2015,canneson2017,zhang2018}, coherent spin dynamics~\cite{Odenthal2017,belykh2019,garcia-arellano2021,kirstein2021}, and nuclear magnetic resonance~\cite{sharma1987,hanrahan2018,aebli2020}. 
 
Here, we study the Land\'e factors of electrons and holes for representative crystals out of the class of lead halide perovskites: MAPbI$_3$, \add{MAPb(Br$_{0.5}$Cl$_{0.5}$)$_3$, MAPb(Br$_{0.05}$Cl$_{0.95}$)$_3$}, FA$_{0.9}$Cs$_{0.1}$PbI$_{2.8}$Br$_{0.2}$\add{, FAPbBr$_3$} and CsPbBr$_3$. Pump-probe Kerr rotation and spin-flip Raman scattering with ultimate resolution in the temporal and spectral domains, respectively, are used to measure the Land\'e factor tensor components in magnetic fields \add{ranging from 20~mT} up to 10~T. The discovered universal dependence of the $g$-factors on the band gap energy is confirmed by first-principles DFT calculations combined with tight-binding and $\bm k\cdot \bm p$ perturbation theory. Thereby we get access to the key band structure parameters and develop a reliable model to predict the Land\'e factors for the whole family of hybrid and inorganic lead halide perovskites, both for bulk crystals and  nanostructures.

\begin{figure}[t!] 
\centering
\includegraphics[width=0.99\columnwidth]{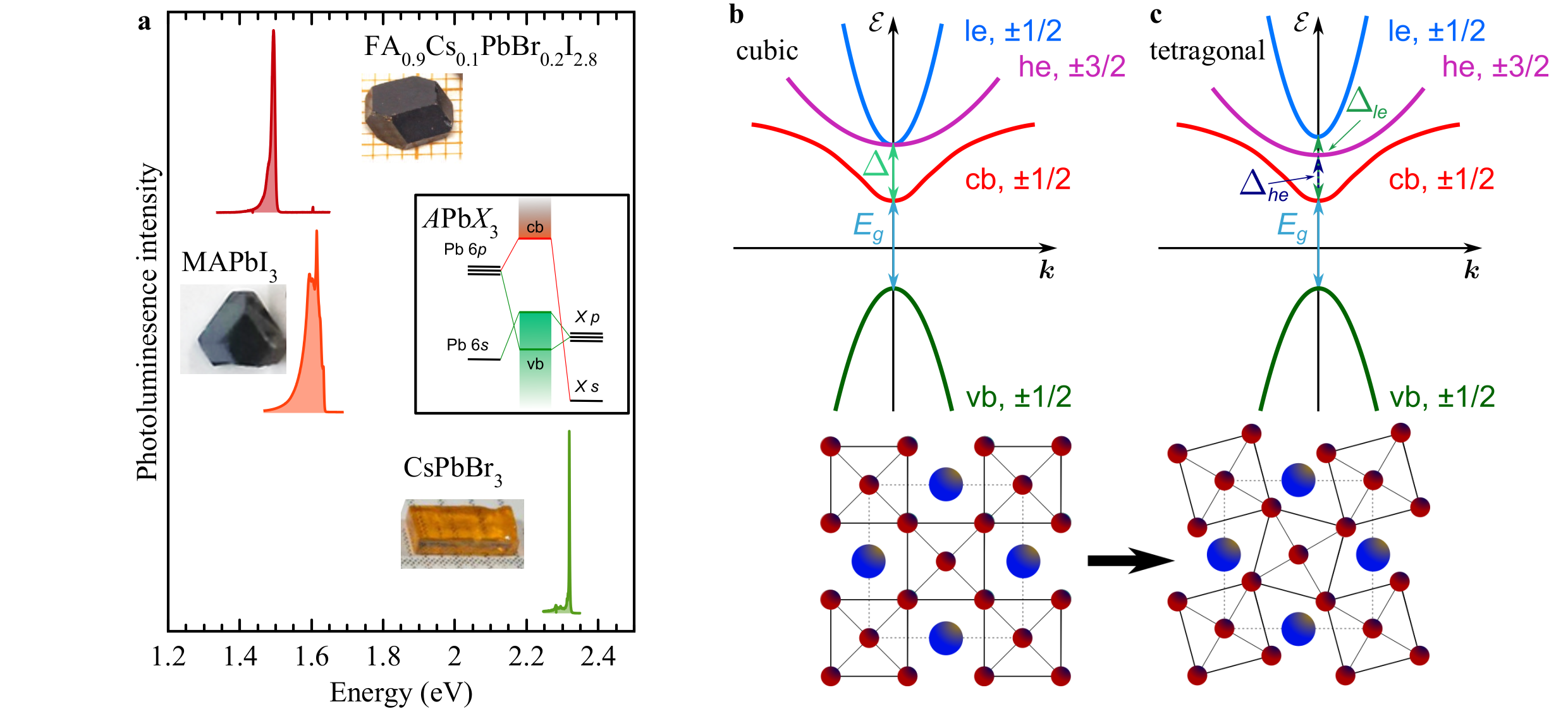}
\caption{\textbf{Lead halide perovskite samples.}
\textbf{a}, Photoluminesence spectra and photographs of the crystals under study. The PL is measured at $T=5$~K. Inset shows a sketch of the contribution of atomic orbitals to the electronic bands around the band gap in lead halide perovskites. \textbf{b}, \textbf{c}, Band structure around the direct gap for the cubic and tetragonal lattices. The states at the top of the valence band (vb, holes) and the bottom of the conduction band (cb, electrons) are spin-degenerate having spin $\pm 1/2$. The degenerate heavy (he, $\pm 3/2$) and light (le, $\pm 1/2$) electron bands are split-off significantly from the cb by the spin-orbit coupling $\Delta$. A symmetry reduction from cubic to tetragonal lifts the he-le degeneracy, leading to individual split-off energies $\Delta_{he/le}$. In the bottom panels the corresponding lattice structure is sketched with the [Pb$X_6$]$^{-4}$ octahedra (dark red symbols show the halogen $X$ anion) and $A$ the inorganic/organic cation (dark blue). An octahedra twist provides the  phase transition from cubic to tetragonal symmetry for the aristotype cubic structure (dashed line square) \add{\cite{Steele2020,yang2020}}.}
\label{fig:intro}
\end{figure}

\section{Experimental Results}

\add{Out of t}he \add{six} investigated samples \add{three representatives, namely the hybrid organic-inorganic MAPbI$_3$ and FA$_{0.9}$Cs$_{0.1}$PbI$_{2.8}$Br$_{0.2}$ as well as the all-inorganic CsPbBr$_3$ lead halide perovskite are chosen for discussion in detail}, as shown in Fig.~\ref{fig:intro}a, together with photoluminescence (PL) spectra measured at the temperature of $T=5$~K. The PL is contributed by recombination of bound excitons. The band gaps of the different materials vary from 1.527~eV up to 2.352~eV, resulting in the black and yellow colors of the studied crystals. 

A sketch of the atom orbitals contributing to the electronic bands in the band gap vicinity is shown in the insert of Fig.~\ref{fig:intro}a. The valence band (vb) is formed by hybridization of Pb $s$ and halogen $X$ $p$ orbitals and the conduction band by Pb $p$ orbitals with some hybridization with the halogen $X$ $s$ orbital.  The band structure  around the direct band gap for cubic crystal symmetry is shown in Fig.~\ref{fig:intro}b, see SI for details. The valence band is simple and has spin $\pm 1/2$. The lowest conduction band (cb) has also spin $\pm 1/2$, while the states of the heavy (he, $\pm 3/2$) and light (le, $\pm 1/2$) electrons are split from the cb by the spin orbit coupling $\Delta$. Commonly, perovskite crystals have cubic crystal symmetry at elevated temperatures, but at lower temperatures they undergo a phase transition by octahedral tilting (see the sketch in the bottom panel of Fig.~\ref{fig:intro}c) \add{to the tetragonal phase and, with further temperature reduction, to the orthorhomic phase}~\cite{Steele2020}. \add{In a nutshell, the condensation of M and R zone-boundary phonons with decreasing temperature \cite{trendel1982,fujii1974} results in reconfiguration of chemical bonds such that the densest package configuration is realized \cite{Steele2020,yang2020}}. \add{The octahedral tilting} lifts the degeneracy of the light and heavy electron states, $\Delta_{le}>\Delta_{he}$, see the upper part of Fig.~\ref{fig:intro}c. 

The Zeeman splitting of the carrier spin states in magnetic field ${\bm B}$ is described by the Hamiltonian
\begin{equation}
\label{Zeeman:gen:1}
\mathcal H_{Z} = \frac{\mu_B}{2} g_{\alpha\beta}  \sigma_\alpha B_\beta,
\end{equation}
where $\mu_B$ is the Bohr magneton, the indices $\alpha,\beta=x,y,z$ denote the Cartesian components, $g_{\alpha\beta}$ are the elements of the $g$-factor tensor, and $\sigma_\alpha$ are the spin Pauli matrices. 
The $g$-factor anisotropy \add{can reveal} information on the crystal symmetry \add{(orthorhombic in our case)}, while the \add{magnitudes} of $g_{\alpha\beta}$ are directly linked to the key band structure parameters: the band gaps and the interband momentum matrix elements~\cite{PhysRev.114.90,ivchenko2005}. 

Experimentally, the $g$-factor can be evaluated from the measured Zeeman splitting $E_Z$ by means of
\begin{equation}
\label{Zeeman:splitting}
E_{Z} =  g \mu_B B.
\end{equation}
The Zeeman splitting can be measured by various techniques. In the spectral domain, spin-flip Raman scattering (SFRS) provides the required high resolution~\cite{hafele_chapter_1991}. Here, $E_{Z}$ is equal to the Raman shift from the exciting laser line (Methods). In the temporal domain, time-resolved pump-probe Kerr rotation (TRKR)~\cite{Yakovlev_Ch6} can give access to the coherent spin dynamics of carriers and therefore, the Larmor precession frequency in a transverse magnetic field, which is linked to the $g$-factor via 
\begin{equation}
\label{Larmor}
\omega_{\rm L} = \frac{\mu_B}{\hbar} g B.
\end{equation}
Both techniques are well established in experiments addressing the spin physics in  semiconductors~\cite{ivchenko2005,glazov2018}. SFRS is applicable only in strong magnetic fields exceeding a few Tesla to obtain a sufficiently large shift from the laser for detection, but is to be preferable for identification of the involved electronic states and mechanisms via polarization analysis, while TRKR has a high precision even in weak magnetic fields and gives access to the spin dynamics. We combine the strengths of both techniques to measure the values and anisotropies of the electron and hole $g$-factors.        

\begin{figure}
\includegraphics[width=0.99\textwidth]{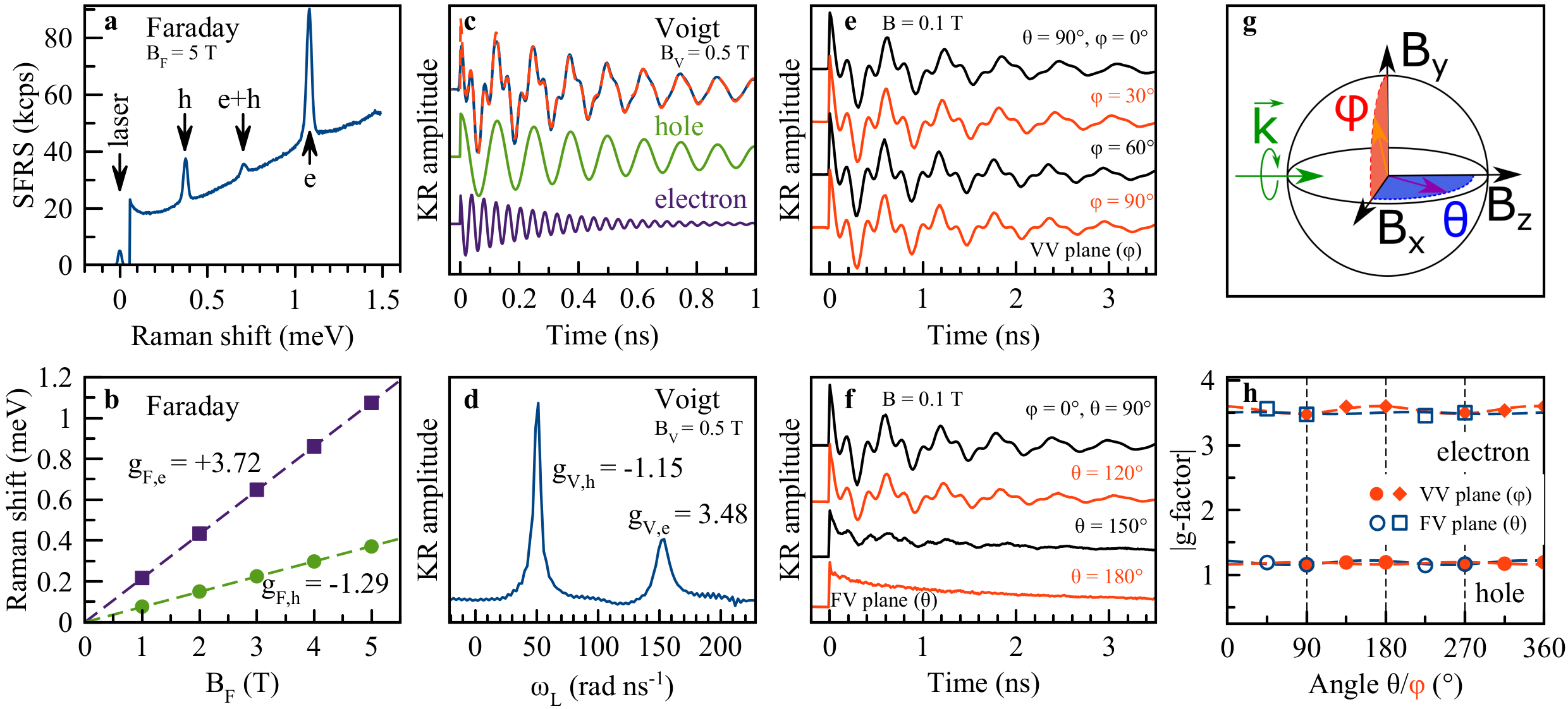}
\caption{\textbf{Pump-probe Kerr rotation and spin-flip Raman scattering of} \textbf{FA}$_\mathbf{0.9}$\textbf{Cs}$_\mathbf{0.1}$\textbf{PbI}$_\mathbf{2.8}$\textbf{Br}$_\mathbf{0.2}$. 
\textbf{a}, SFRS spectrum in Faraday geometry ($\theta=0^\circ$, $\varphi=0^\circ$) measured at $T=1.6$~K. The excitation/detection polarization is $\sigma^+$/$\sigma^-$.  
\textbf{b}, Magnetic field dependences of the measured Raman shifts (symbols).  Lines are linear fits.
\textbf{c}, TRKR signal (blue line) measured in the Voigt geometry ($\theta=90^\circ$, $\varphi=0^\circ$). Red dashed line is fit with two oscillatory functions (see Methods) with the parameters: $S_{\rm e}/S_{\rm h}=0.8$, $T^*_{\rm 2,e}(B=0.5~\textrm{T})=275$~ps, $T^*_{\rm 2,h}(B=0.5~\textrm{T})=690$~ps, $g_h=-1.15$ and $g_e=+3.48$. Decomposed electron and hole components are given below. TRKR data are measured at $T=5$~K. \textbf{d}, Fast Fourier transform spectrum of the signal from panel \textbf{c}.  
\textbf{e}, \textbf{f}, TRKR signals in magnetic fields of 0.1~T rotated in the Voigt-Voigt and Faraday-Voigt planes, respectively. \textbf{g}, Laboratory coordinate system: blue  horizontal (red vertical) plane is the Faraday-Voigt (Voigt-Voigt) plane. \textbf{h}, Angular dependences of the electron and hole $g$-factors for the magnetic field rotated in the Faraday-Voigt (open symbols) and Voigt-Voigt (closed symbols) planes. Note that while $g_{\rm h}<0$ we show $|g_{\rm h}|$.  Lines are guides to the eye.  The indices V and F indicate the Voigt and Faraday directions of the magnetic field.} 
\label{fig:FAPI}
\end{figure}

In presenting the experimental data, let us start with the FA$_{0.9}$Cs$_{0.1}$PbBr$_{0.2}$I$_{2.8}$ crystal. In the corresponding SFRS spectrum measured in the Faraday geometry at $B_{\rm F}=5$~T, pronounced lines associated with the electron and hole spin-flips are detected, see Fig.~\ref{fig:FAPI}a. The larger shift of the electron line corresponds to a larger Zeeman splitting, i.e. to a larger absolute value of the electron $g$-factor compared to the hole $g$-factor. Analysis of the polarization properties in the Faraday and Voigt geometries allows us to conclude that the SFRS signals are provided by the spin-flip of a resident electron or hole, interacting with a photogenerated exciton~\cite{Debus2013}. The resident carriers are created by photogeneration, and are localized at separate crystal sites for cryogenic temperatures~\cite{belykh2019}. The specific feature of lead halide perovskites is the coexistence of resident electrons and holes, which is unusual for common semiconductors. Here, we make use of this feature, as in one crystal both the electron and hole properties can be studied in the same experiment. The e+h SFRS line in Fig.~\ref{fig:FAPI}a has a shift corresponding to  $g_{\mathrm e} + g_{\mathrm h}$, evidencing that it is due to a combined spin-flip of an electron and a hole, interacting with the same exciton. From the linear dependence of the Raman shifts on magnetic field, using  equation~\eqref{Zeeman:splitting}, we evaluate $g_{\rm F,h}=-1.29$ and $g_{\rm F,e}=+3.72$, Fig.~\ref{fig:FAPI}b. Note that both shift dependencies show no offset at zero magnetic field,  confirming the involvement of a resident carrier, as a finite shift would be expected due to electron-hole exchange interaction for carriers bound within an exciton~\cite{ivchenko2005}. \add{Further we do not observe any deviations from the linear law~[Eq.\eqref{Zeeman:splitting}] even at the smallest applied magnetic fields showing that the Rashba effect is negligible.}

In pump-probe Kerr rotation (KR) experiments, carrier spin polarization along the light wave vector direction $\textbf{k}$ is induced by the circularly polarized pump pulses, Fig.~\ref{fig:FAPI}g and Methods. The spin polarization dynamics are detected via the Kerr rotation of the linearly-polarized probe pulses. Commonly, KR is measured in Voigt geometry where the Larmor spin precession about the magnetic field leads to an oscillating decaying signal. An example for $B_{\rm V}=0.5$~T is shown in Fig.~\ref{fig:FAPI}c. The signal is contributed by two frequencies corresponding to $g_{\rm h}=-1.21$ and $g_{\rm e}=+3.57$, as can be seen from the Fast Fourier Transformation (FFT) spectrum in Fig.~\ref{fig:FAPI}d and also from fitting it with two oscillating functions (Methods). The decomposed spin dynamics of holes and electrons are shown in Fig.~\ref{fig:FAPI}c.      

The $g$-factor anisotropy is measured by tilting the magnetic field in a vector magnet. The field direction is defined by two angles:  $\theta$ for rotation in the Faraday-Voigt (FV) plane (blue) and $\varphi$ for rotation in the Voigt-Voigt (VV) plane (red), see Fig.~\ref{fig:FAPI}g. For VV plane rotation ($\textbf{B} \perp \textbf{k}$) the TRKR signals are very similar, see Fig.~\ref{fig:FAPI}e. For rotation from Voigt to Faraday geometry (FV plane), see Fig.~\ref{fig:FAPI}f, the spin precession amplitude decreases while the amplitude of the monotonically decaying signal increases. The latter corresponds to the signal in Faraday geometry ($\theta=0^\circ$ or $180^\circ$) reflecting the longitudinal spin relaxation. The $g$-factors evaluated from Figs.~\ref{fig:FAPI}e,f are collected in Fig.~\ref{fig:FAPI}h. One can see that both the electron and hole $g$-factors are pretty isotropic in the FA$_{0.9}$Cs$_{0.1}$PbI$_{2.8}$Br$_{0.2}$ crystals, varying in the ranges from $+3.48$ to $+3.60$ and from $-1.15$ to $-1.22$, respectively.

\begin{figure}
\includegraphics[width=1\textwidth]{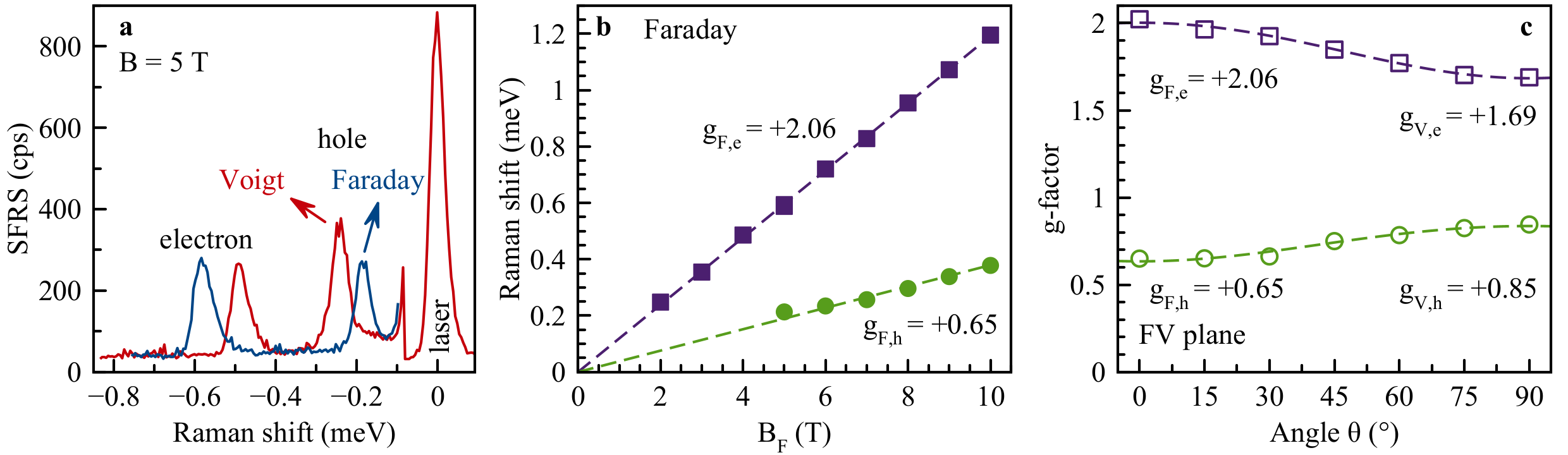}
\caption{\textbf{Spin-flip Raman scattering of CsPbBr$_3$}. 
\textbf{a}, SFRS spectra measured in the Faraday (blue) and Voigt (red) geometries at $T=1.6$~K. The excitation/detection polarization is $\sigma^-$/$\sigma^+$. Spectra are shown for the anti-Stokes spectral range where photoluminescence is suppressed. 
\textbf{b}, Magnetic field dependence of the electron (purple squares) and hole (green circles) Stokes Raman shifts.  Lines are linear fits with $g_{\rm F,e}=+2.06$ and $g_{\rm F,h}=+0.65$. 
\textbf{c}, Angular dependences of the electron and hole $g$-factors for the magnetic field rotated in the Faraday-Voigt plane  ($\varphi=0^\circ$). Lines are fits using equation~(\ref{eq:aniso}). The sample $c$-axis is oriented perpendicular to $\textbf{k}$ and parallel to the x-axis, $\textbf{c} \parallel \textbf{B}$, corresponding to the Voigt geometry ($\theta=90^\circ$).
}
\label{fig:CsPBr}
\end{figure}

Let us turn to the CsPbBr$_3$ crystal with the $c$-axis perpendicular to $\textbf{k}$. SFRS spectra measured in the Faraday and Voigt geometries are shown in  Fig.~\ref{fig:CsPBr}a. They reveal a difference in the Raman shifts evidencing a pronounced anisotropy of the carrier $g$-factors: $g_{\rm F,e}=+2.06$, $g_{\rm V,e}=+1.69$ for the electrons and $g_{\rm F,h}=+0.65$,  $g_{\rm V,h}=+0.85$ for the holes. The magnetic field dependence of the Raman shift presented in Fig.~\ref{fig:CsPBr}b together with the TRKR results for this sample published in Ref.~\cite{belykh2019} allows us to conclude that in CsPbBr$_3$, similarly to FA$_{0.9}$Cs$_{0.1}$PbBr$_{0.2}$I$_{2.8}$, the spin signals are contributed by resident carriers. The measured anisotropy in the FV plane is given in Fig.~\ref{fig:CsPBr}c. It can be well described by 
\begin{equation}
\label{eq:aniso}
g_{\rm e(h)}(\theta,\varphi=0^\circ)=\sqrt{g^2_{\rm F, e(h)} \cos^2 \theta+g^2_{\rm V, e(h)} \sin^2 \theta} .
\end{equation}

\FloatBarrier

\begin{figure}[ht!] 
		\includegraphics[width=1\columnwidth]{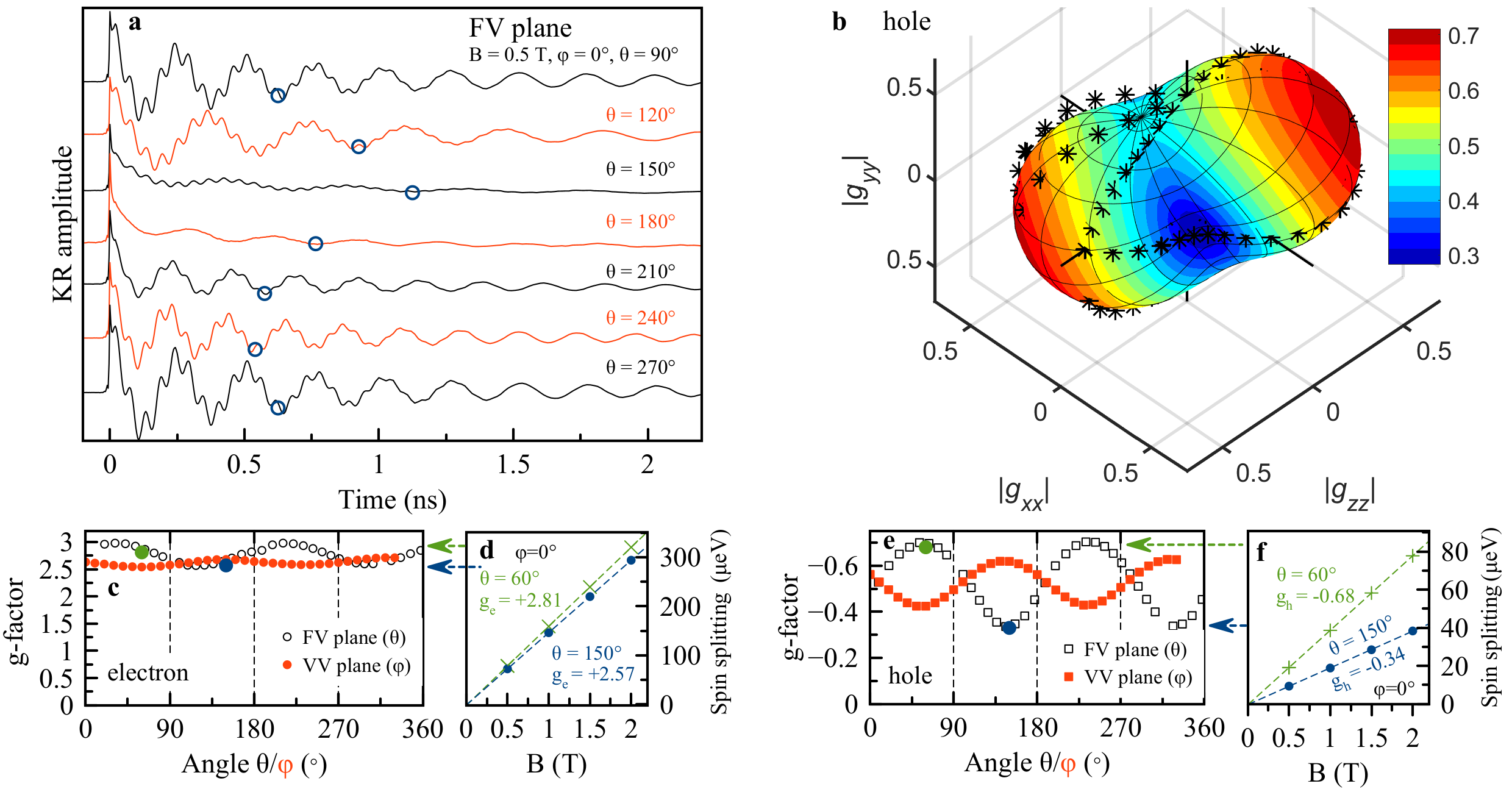} 
\caption{\textbf{Pump-probe Kerr rotation in MAPbI$_3$.} 
\textbf{a}, TRKR signals in magnetic fields of 0.5~T rotated in the Faraday-Voigt plane ($\varphi=0^\circ$). $T=7$~K. The variation of the hole Larmor frequency is indicated by the open circles. \textbf{b}, Three-dimensional presentation of the hole $g$-factor tensor. Stars are experimentally measured values. The contour is a fit to the data using \ref{eq:aniso} \textbf{c,e}, Angular dependencies of the electron and hole $g$-factors for the magnetic field rotated in the Faraday-Voigt (open symbols) and Voigt-Voigt (closed symbols) planes. \textbf{d,f}, Magnetic field dependencies of the electron and hole Zeeman splittings evaluated from their Larmor frequencies. Experimental data (symbols) are measured at $\theta=60^\circ$ and $150^\circ$ ($\varphi=0^\circ$). Lines are linear fits.} \label{fig:MAPI}
\end{figure}

In the MAPbI$_3$ crystal, as well as in the two other materials, the spin precession of resident electrons and holes can be also well resolved in the TRKR signals, Fig.~\ref{fig:MAPI}a. For rotation of the field orientation in the FV plane, a strong anisotropy of the hole $g$-factor can be concluded from the variation of the hole precession period, e.g., following the third minimum of the hole precession as indicated by the circles. The electron $g$-factor is also anisotropic. The $g$-factor variations for rotation in the FV and VV planes are given in  Figs.~\ref{fig:MAPI}c,e. For the holes it ranges between $-0.28$ and $-0.71$, while for the electrons is varies between $+2.46$ and $+2.98$. Interestingly, the $g$-factor extremal values do not coincide with the main cubic axes given by the angles $\theta, \varphi=\{0^\circ,90^\circ,180^\circ,270^\circ\}$. We have recorded a large data set by measurements for further angles not falling into the FV and VV planes and found that the main axis of the electron $g$-tensor has the orientation $\theta=33^\circ$, $\varphi=54^\circ$ and for the hole $g$-tensor its direction is $\theta=57^\circ$, $\varphi=54^\circ$. The hole $g$-tensor is visualized as three-dimensional plot in Fig.~\ref{fig:MAPI}b, for the electron see the SI. SFRS measurements are also working well for the MAPbI$_3$ crystal, yielding the same $g$-factor values as we show in SI.

\section{Model and discussion}

\begin{table}[t]
\caption{Electron and hole $g$ factors measured in the studied perovskite crystals at $T=1.6$ and 5~K. Minimum and maximum values are presented together with the degree of anisotropy defined as $P_{\rm e(h)}=100\% \times (g_{\rm max}-g_{\rm min})/(g_{\rm max}+g_{\rm min})$.}

\label{tab:gfactors}
\begin{tabular}{|c|c|c|c|c|c|c|}
\hline
Material &  $E_g$~(eV) & $g_{\rm e}$ & $P_{\rm e}$ &$g_{\rm h}$ & $P_{\rm h}$  & Comments  \\
\hline
FA$_{0.9}$Cs$_{0.1}$PbI$_{2.8}$Br$_{0.2}$ & 1.527 & +3.48 to +3.60  & 2\% & $-1.15$ to $-1.22$ & $<$4\%  & isotropic \\
MAPbI$_3$   & 1.652 \cite{galkowski2016}  & +2.46 to +2.98 & 10\% &  $-0.28$ to  $-0.71$  & 43\% & anisotropic, tilted\\
FAPbBr$_{3}$& 2.189  & +2.32 to +2.44 &  & +0.36 to +0.41 &  & isotropic\\
CsPbBr$_3$  & 2.352 \cite{belykh2019}   & +1.69 to +2.06 & 10\%  & +0.65 to +0.85 & 13\%   & anisotropic\\ 
MAPb(Br$_{0.5}$Cl$_{0.5}$)$_{3}$   & 2.592  & +1.47 &  &    &  & SFRS - z axis\\
MAPb(Br$_{0.05}$Cl$_{0.95}$)$_{3}$   & 3.157 &  &  & +1.33 &  & TRKR - x axis\\
\hline
\end{tabular}
\end{table} 

\begin{figure}[ht]
		\includegraphics[width=1\columnwidth]{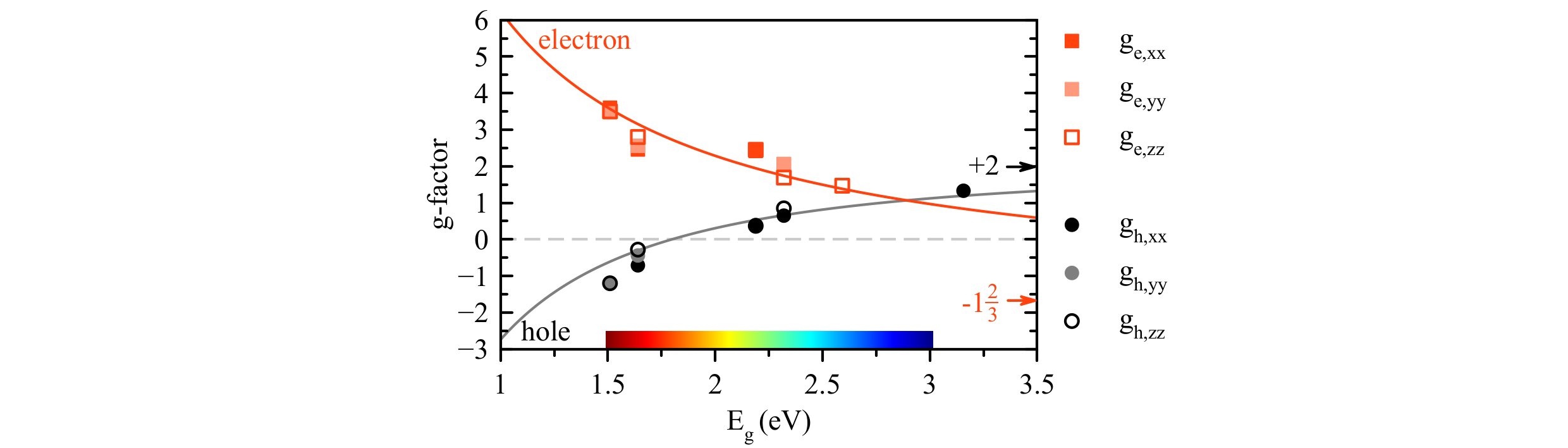}
\caption{\textbf{Electron and hole $g$-factors versus band gap energy in lead halide perovskite crystals.} Experimental data are given by the symbols. Solid lines are fits with Eqs.~\eqref{gv:cub:1} and \eqref{gc:cub:1} using the parameters $\Delta=1.5$~eV, $\hbar p/m_0 = 6.8$~eV{\AA}, and $\Delta g_e =-1$. The limiting values of the electron ($-5/2$) and hole ($+2$) $g$-factors for $E_g \to \infty$ are given by the arrows.  
}
\label{fig:summary}
\end{figure}

Our experimental results for the Land\'e factors of electrons and holes are summarized in the Table~\ref{tab:gfactors} and plotted in Fig.~\ref{fig:summary}, where the $g$-factor tensor components are shown as function of the band gap. The clear correlation of the $g$-factors and the band gap energy requires a theoretical explanation. \add{Note, that we do not observe any deviations from the linear Zeeman term [Eq. (\ref{Zeeman:splitting})] even in the smallest applied magnetic fields, hence we neglect the Rashba term in the analysis.} To that end we use atomistic approaches based on the density functional theory (DFT) and empirical tight-binding method (ETB), to calculate the band structure and Land\'e factors of prototypical inorganic analogues of the studied perovskites: CsPbI$_3$, CsPbBr$_3$, and CsPbCl$_3$~\add{\cite{Nestoklon:2021wg}}. Although these all-inorganic systems are different from organic-inorganic lead halides, resulting in somewhat different band gaps, spin-orbit couplings and Land\'e factors, the organic molecules are not crucial for the band structure formation, so that the study of the inorganic perovskites allows us to establish general trends and, eventually, formulate a semi-phenomenological model for the Zeeman splitting.  

The details of the atomistic calculations and the $\bm k\cdot\bm p$ model description of the band structure are given in the SI. We express the components of the $g$-factor tensors via the matrix elements of the spin and orbital angular momenta of electrons and holes which can be further expressed via the interband momentum matrix elements and band gaps. The analysis of the influence from the different bands allows us to identify the key contributions to the electron and hole Land\'e factors. In particular, for the holes in the valence band the main contribution is related to the $\bm k\cdot\bm p$ mixing with the conduction band \add{\cite{PhysRev.114.90,ivchenko2005,yu2016}} and, in the cubic phase one finds~\add{(see SI for details)}
\begin{equation}
\label{gv:cub:1}
g_{h} = 2- \frac43 {\frac{p^2}{m_0}}  \left(\frac{1}{E_g} -\frac{1}{E_g+\Delta}\right).
\end{equation}
Here $p$ is the interband matrix element of the momentum operator, $E_g$ the band gap and $\Delta$ the spin-orbit splitting of the conduction band. For the electrons, the $\bm k\cdot\bm p$ mixing both with the top valence band and the remote valence states is important, resulting in~\add{(cf. Ref~\cite{yu2016} and SI)}
\begin{equation}
\label{gc:cub:1}
g_{e} = -\frac{2}{3}+ \frac43 \frac{p^2}{m_0E_g}  + \Delta g_e,
\end{equation}
where $\Delta g_e$ is the remote band contribution. Equations~\eqref{gv:cub:1} and \eqref{gc:cub:1} with the reasonably chosen parameters $\Delta=1.5$~eV, $\hbar p/m_0 = 6.8$~eV$\cdot$\AA, and $\Delta g_e =-1$  describe well the band gap dependence of the Land\'e factors across the studied perovskite classes, see Fig.~\ref{fig:summary}. In agreement with experiment, the $\bm k\cdot\bm p$-mixing between the conduction and valence bands decreases with increasing band gap, thus the electron $g$-factor decreases with increasing $E_g$ from large positive values down to $-2/3+\Delta g_e \approx - 5/3$, while the hole $g$-factor increases with increasing $E_g$ from negative values up to $+2$. Note that in the studied range of band gap energies the separation from the remote bands is large, so that $\Delta g_e$ is practically independent of the material. Also the spin-orbit interaction is about constant since it is determined by the heavy lead atoms. 

Let us now address the anisotropy of the $g$-factors observed mainly in the experiments on MAPbI$_3$ and CsPbBr$_3$. Overall, the $g$-factor anisotropy may be expected because at low temperatures the perovskites are known to undergo phase transitions from cubic to tetragonal or orthorhombic phases, see~\cite{Steele2020}. In the simplest approach the tetragonal and orthorhombic phases can be considered as a cubic phase distorted along the principal cubic axes $x$, $y$, and $z$. In the tetragonal phase $g_{xx} = g_{yy} \ne g_{zz}$, where $z$ is the $C_4$ axis. In the orthorhombic phase, on the other hand, $g_{xx} \ne g_{yy} \ne g_{zz}$. The analysis shows (see SI) that the $g$-factor anisotropy is caused by the conduction band crystal splitting with $\Delta_{he} \ne \Delta_{le}$, see Fig.~\ref{fig:intro}\add{c}, and by the anisotropy of the interband momentum matrix elements. The conduction band splitting affects mainly the hole $g$-factor [Eq.~\eqref{gv:cub:1}], while the anisotropy of matrix elements affects both the electron and hole Land\'e factors, see the SI where details of the fits are presented. This, together with the fact that $|g_h| < g_e$, results in a larger anisotropy of the hole $g$-factors, $P_h>P_e$, as observed in the experiment.  

For the cubic, tetragonal and orthorhombic symmetries the main axes of the $g$-factor tensor are related to the cubic directions $\langle 001 \rangle$. We confirm this experimentally for CsPbBr$_3$, \add{FA$_{0.9}$Cs$_{0.1}$PbI$_{2.8}$Br$_{0.2}$ and FAPbBr$_3$}, while in MAPbI$_3$ the main axes are found to be tilted with respect to the cubic axes.  This might be a signature of the previously not reported transition to a monoclinic phase or a rotation of the crystallographic axes with respect to the lab frame \cite{onoda1990,swainson2003,whitfield_structures_2016}. In the former case,  the reduced symmetry of the structure can be described by one of two possible point groups (both being subgroups of $D_{2h}$): $C_1$ (no non-trivial symmetry operations) or $C_i$ (with space inversion) as symmetry operation. All other subgroups imply that at least one of the main axes of the $g$-factor tensor is the ``cubic'' axis $[001]$, $[010]$ or $[100]$. \add{Alternatively, the explanation could be related to the formation or randomly oriented domains with different crystalline orientations.} This question requires further experimental studies, particularly because X-ray diffraction for determination of the crystal structure at low temperatures is quite involved. 

\section{Conclusion}

We have discovered experimentally and confirmed theoretically an universal dependence of the electron and hole $g$-factors on the band gap energy in the family of lead halide perovskite materials which is applicable for hybrid organic-inorganic and all-inorganic compounds.  Our first-principles DFT calculations in combination with tight-binding and $\bm k\cdot \bm p$ approaches show that the universality originates from the electronic states that form the band gap being largely contributed by lead orbitals. Therefore, the $g$-factor dependence across  the huge range of band gap energies from 1~eV up to 4~eV can be treated with the same band parameters. The derived parameters give insight into fundamental band properties, including band anisotropies caused by structural phase transitions. The selection of several halogen atoms (I, Br, Cl) and cation types (Cs, MA, FA) allows tailoring of the band gaps, which in turn leads to a correpsonding variation of the $g$-factor values. Our study provides a reliable relation to predict the Land\'e factors for the whole family of lead halide perovskites, including their nanostructures. This relation thus delivers the key parameter determining the spin physics of perovskites.

\section*{Methods}

\textbf{Samples.}
The class of lead halide perovskites possesses $A$Pb$X_3$ composition, where the $A$-cation is typically Cs, methylammonium (MA, CH$_3$NH$_3$) or formamidinium (FA, CH(NH$_2$)$_2$) and the $X$-anion is a halogen Cl, Br, or I, giving rise of a huge flexibility. The latter is only limited by a favorable ratio of the anion to cation ion radii, named the Goldschmidt tolerance factor $t$, which should be close to unity~\cite{goldschmidt_gesetze_1926}. By varying composition, the band gap of these perovskite materials can be tuned from the infrared up to the ultraviolet spectral range. All studied samples are lead halide perovskite single crystals grown out of solution with the inverse temperature crystallization (ITC) technique~\cite{Dirin2016,nazarenko2017,hocker2021}. For the specific crystals the ITC protocols were modified.     
 
\textbf{FA$_{0.9}$Cs$_{0.1}$PbI$_{2.8}$Br$_{0.2}$ crystals.}
$\alpha$-phase FA$_{0.9}$Cs$_{0.1}$PbI$_{2.8}$Br$_{0.2}$ single crystals were grown as reported in Ref.~[\onlinecite{nazarenko2017}]. First, a solution of CsI, FAI (FA being formamidinium), PbI$_2$, and PbBr$_2$, with GBL $\gamma$-butyrolactone as solvent is mixed. This solution is then filtered and slowly heated to 130$^\circ$C temperature, whereby the single crystals are formed in the black phase of FA$_{0.9}$Cs$_{0.1}$PbI$_{2.8}$Br$_{0.2}$. Afterwards the crystals are separated by filtering and drying. The $\alpha$-phase (black phase) exhibits a cubic crystal structure at room temperature~\cite{weller_cubic_2015}. In the experiment the crystal was oriented with $[001]$ pointing along the laser wave vector $\textbf{k}$. \add{Note that the $g$-factor isotropy, the small shift of the PL line with temperature and further analyses \cite{nazarenko2017,kirstein2021} suggest the typical lead halide perovskite crystal distortion from cubic symmetry to be small at small temperatures.} The size of the crystal is $\approx2 \times 3\times 2$~mm$^3$. The crystal shape is non-cuboid, but the crystal structure exhibits aristotype cubic symmetry. \add{Sample code: 515a. }

\textbf{CsPbBr$_3$ crystals.} 
The CsPbBr$_3$ crystals were grown with a slight modification of the ITC, see Ref.~\cite{Dirin2016}. First, CsBr and PbBr$_2$ were dissolved in dimethyl sulfoxide. Afterwards a cyclohexanol in N,N-dimethylformamide solution was added. The resulting mixture was heated in an oil bath to $105^\circ$C whereby slow crystal growth appears. The obtained crystals were taken out of the solution and quickly loaded into a vessel with hot ($100^\circ$C) N,N-dimethylformamide. Once loaded, the vessel was slowly cooled down to about $50^\circ$C. After that, the crystals were isolated, wiped with filter paper and dried. The obtained rectangular-shaped CsPbBr$_3$ is crystallized in the orthorhombic modification. The crystals have one selected (long) direction along the $c$-axis [002] and two nearly identical directions along the $[\bar{1}10]$ and [110] axes~\cite{feng2020}. The size of the crystal is $\approx3\times2\times7$~mm$^3$. \add{Sample code: DD4470/2}

\textbf{MAPbI$_3$ crystals.}
Methylammonium (MA/CH$_3$NH$_3$) lead tri-iodine (MAPbI$_3$) single crystals were low temperature solution-grown in a reactive inverse temperature crystallization (RITC) process, Ref.~\cite{hocker2021}, which utilizes a mixture of $\gamma$-butyrolactone GBL precursor solvent with alcohol. The mixed precursor solvent polarity is changed compared to pure GBL, causing a lower solubility of MAPbI$_3$ and an optimization of nucleation rates and centers, which result in an early crystallization at low temperatures. Black MAPbI$_3$ single crystals were obtained at a temperature of 85$^\circ$C. At room temperature a tetragonal phase with lattice constants $a=0.893$~nm and $c=1.25$~nm was determined by XRD~\cite{hocker2021}. The size of the crystal is $\approx4 \times 3\times 2$~mm$^3$. The crystal shape is non-cuboid, but the crystal structure exhibits aristotype cubic symmetry. The front facet was X-ray characterized to point along the a-axis \cite{hocker2021}. \add{Sample code: MAPI-SC04}

\add{\textbf{MAPb(Br$_{0.5}$Cl$_{0.5}$)$_3$ \& MAPb(Br$_{0.05}$Cl$_{0.95}$)$_3$ crystals.}}
\add{In the case of MAPb(Br$_{0.5}$Cl$_{0.5}$)$_3$, 0.3~mmol MACl, 0.3~mmol PbCl$_2$, 0.7~mmol MABr and 0.7~mmol PbBr$_2$ were dissolved in a mixture of 0.89~ml dimethylformamide (DMF) with 85~$\mu$l dimethyl sulfoxide (DMSO). In the case of MAPb(Br$_{0.05}$Cl$_{0.95}$)$_3$, 2.2~mmol PbCl$_2$, 1.2~mmol MACl and 1~mmol MABr were dissolved in 2~ml of 1:1 DMF:DMSO mixture. Both solutions were filtered through 0.45~$\mu$m polytetrafluoroethylene (PTFE) filter. The obtained solutions were slowly heated in an oil bath up to 62$^\circ$C. The crystals nucleate and grow in the temperature window 58--62$^\circ$C. The obtained crystals show a rectangular cuboid shape with sizes of $1.64 \times 1.65 \times 2.33$~mm$^3$ for MAPb(Br$_{0.05}$Cl$_{0.95}$)$_3$ and $2 \times 4 \times 1$~mm$^3$ for MAPb(Br$_{0.5}$Cl$_{0.5}$)$_3$. The MAPb(Br$_{0.05}$Cl$_{0.95}$)$_3$ crystals are transparent and colorless, while the MAPb(Br$_{0.5}$Cl$_{0.5}$)$_3$ crystals are transparent with a faint yellow appearance. Sample code: dd2924 and dd6347, respectively.}

\add{\textbf{FAPbBr$_3$ crystals.}}
\add{The FAPbBr$_3$ single crystals were grown with an analogous approach, following \cite{saidaminov2015}. The crystal is of reddish transparent appearance and shows a rectangular cuboid shape with a size of $5 \times 5 \times 2$~mm$^3$. Sample code: OH0071a.}

\textbf{Magneto-optical measurements.} 
The samples were placed in a cryostat with the temperature variable from 1.6~K up to 300~K. For $T=1.6$~K the sample is immersed in superfluid helium while for 4.2~K to 300~K the sample is in cooling helium gas. Two magnet cryostats equipped with split-coil superconducting solenoids were used. The first one is constructed to generate magnetic fields up to 10~T in a fixed direction. The second one, which is a vector magnet, has three pairs of orthogonal split coils, allowing us to apply magnetic fields up to 3~T in any direction. A sketch of the experimental geometry is shown in Fig.~\ref{fig:FAPI}g. A magnetic field parallel to the light wave vector \textbf{k} is denoted here as $\textbf{B}_z$ (Faraday geometry), magnetic fields perpendicular to \textbf{k} (Voigt geometry) are oriented in the plane spanned by the magnetic field axis $B_x$ in the horizontal plane and $B_y$ in the vertical direction. The angle $\theta$ is defined as the angle between $\textbf{B}_x$ and $\textbf{B}_z$ (for rotation in the Faraday-Voigt (FV) plane) with $\theta$=0$^\circ$ corresponding to $\textbf{B}_z \parallel \textbf{k}$. The angle $\varphi$ defines the rotational orientation in the vertical plane (Voigt-Voigt (VV) plane) with $\varphi=0^\circ$ for the horizontal x-axis and $\varphi=90^\circ$ for the vertical y-axis. 

\textbf{Pump-probe time-resolved Kerr rotation (TRKR).}
The coherent spin dynamics were measured by a pump-probe setup, where pump and probe had the same photon energy, emitted from the same pulsed laser~\cite{Yakovlev_Ch6}. A titan-sapphire (Ti:Sa) laser emitted 1.5~ps long pulses with a spectral width of about $1$~nm (1.5~meV) at a pulse repetition rate of 76~MHz (repetition period $T_\text{R}=13.2$~ns). The laser photon energy was tunable in the spectral range of $1.265-1.771$~eV ($700 - 980$~nm). It was set to the vicinity of the exciton resonance at the maximum of the Kerr rotation signal: at 1.513~eV for FA$_{0.9}$Cs$_{0.1}$PbI$_{2.8}$Br$_{0.2}$ and at 1.637~eV for MAPbI$_3$. The laser beam was split into two beams (pump and probe). The probe pulses were delayed with respect to the pump pulses by a mechanical delay line.  Both pump and probe beams were modulated using photo-elastic modulators (PEM). The probe beam was always linearly polarized and its amplitude was modulated at a frequency of 84~kHz. The pump beam helicity was modulated between $\sigma^+$ and $\sigma^-$ circular polarization at a frequency of 50~kHz. The polarization of the reflected probe beam was analyzed, via a lock-in technique, with respect to the rotation of its linear polarization (Kerr rotation). In finite transverse magnetic field, the Kerr rotation amplitude oscillates in time reflecting the Larmor spin precession of the carriers and decays at longer time delays. When both electrons and holes contribute to the Kerr rotation signal, which is the case for the studied perovskite crystals, the signal can be described as a superposition of two decaying oscillatory functions: $A_{\rm KR} = S_{\rm e} \cos (\omega_{\rm L, e} t) \exp(-t/T^*_{\rm 2,e}) + S_{\rm h} \cos (\omega_{\rm L, h} t) \exp(-t/T^*_{\rm 2,h})$. Here $S_{\rm e(h)}$ are the signal amplitudes that are proportional to the spin polarization of electrons (holes). The $g$-factors are evaluated from the Larmor precession frequency $\omega_{\rm L, e(h)}$ by means of $|g_{\rm e(h)}|=  \hbar \omega_{\rm L, e(h)}/ (\mu_{\rm B} B)$. It is also important to note, that TRKR provides information on the $g$-factor magnitude, but not on its sign. The same is true for the SFRS technique below. Information on the sign is obtained from model calculations, e.g. for the studied perovskites $g_{\rm e}>0$ is predicted, see SI. Also knowledge on the exciton $g$-factor, $g_{\rm X}$, and its sign can help to identify the hole $g$-factor sign using $g_{\rm X}=g_{\rm e}+g_{\rm h}$~\cite{belykh2019}. For  standard TRKR measurements, the external magnetic field is applied in the Voigt geometry perpendicular to the light wave vector $\textbf{k}$. For measuring the $g$-factor anisotropy the magnetic field orientation is tuned to various angles $\theta, \varphi$ using the vector magnet, see Fig.~\ref{fig:FAPI}g. 
  
\textbf{Spin-flip Raman scattering (SFRS).} The SFRS technique allows one to measure directly the Zeeman splitting of the electron and hole spins from the spectral shift of the scattered light from the laser photon energy~\cite{hafele_chapter_1991,Debus2013}. The energy shift is provided by the spin-flip of carriers, with the required energy taken from or provided by phonons. The typical shifts do not exceed 1~meV at the magnetic field of 10~T, which demands for the high spectral resolution provided by high-end spectrometers with excellent suppression of scattered laser light. The experiments were performed for samples in contact with pumped liquid helium at $T=1.6$~K. We used resonant excitation in the vicinity of the exciton resonances in order to enhance the SFRS signal, obtained for the following laser photon energies: 1.500~eV for FA$_{0.9}$Cs$_{0.1}$PbI$_{2.8}$Br$_{0.2}$, 2.330~eV for CsPbBr$_3$, and 1.635~eV for MAPbI$_3$. The resonant Raman spectra were measured in the backscattering geometry with the incident laser excitation density between 1 and 5~Wcm$^{-2}$. The scattered light was analyzed by a Jobin-Yvon U1000 double monochromator (1 meter focal length) equipped with a cooled GaAs photomultiplier and conventional photon counting electronics. The used spectral resolution of 0.2~cm$^{-1}$ (0.024~meV)  allowed us to measure the SFRS signals in close vicinity of the laser line for spectral shifts ranging from 0.1 to 3~meV. The spectra were measured in co-polarized ($\sigma^+$/$\sigma^+$) or cross-polarized ($\sigma^+$/$\sigma^-$  and $\sigma^-$/$\sigma^+$) circular polarizations of excitation and detection for the Faraday geometry ($\mathbf{B} \parallel \mathbf{k}$) and in crossed or parallel linear polarizations for the Voigt geometry ($\mathbf{B} \perp \mathbf{k}$), with the definition of the magnetic field axes and angles as stated above in Methods (Magneto-optical measurements). For data presentation we plot the Stokes-shifted lines (i.e. the lines shifted to lower energies from the laser) as positive shifts, while for the anti-Stokes lines negative values are taken. Note that in the anti-Stokes spectra  possible contributions of photoluminescence are absent, as up-conversion at low temperatures is weak, so that spin-flip lines can be clearly identified. On the other hand, for the Stokes spectra larger Raman shifts in stronger magnetic fields can be observed.   

\textbf{Data availability.} 
The data on which the plots within this paper are based and other findings of this study are available from the corresponding author upon justified request. \\
\textbf{Code availability.} 
The code on which the calculations within this paper are based and other findings of this study are available from the corresponding author upon justified request. \\

\textbf{ORCID.} \\
Erik Kirstein:        0000-0002-2549-2115 \\
Dmitri R. Yakovlev:   0000-0001-7349-2745 \\
Evgeny A. Zhukov:     0000-0003-0695-0093 \\
Dennis Kudlacik:      0000-0001-5473-8383\\
Ina V. Kalitukha:     0000-0003-2153-6667\\
Victor F. Sapega:     0000-0003-3944-7443\\
Grigorii S. Dimitriev: 0000-0002-6779-0997\\
Mikhail M. Glazov:    0000-0003-4462-0749 \\  
Marina A. Semina:    	   0000-0003-3796-2329\\
Mikhail M. O. Nestoklon:   0000-0002-0454-342X\\
Eugeniyus L. Ivchenko:    0000-0001-7414-462X\\
Dmitry N. Dirin:    0000-0002-5187-4555\\
Maksym V. Kovalenko:  0000-0002-6396-8938 \\
Andreas Baumann:      0000-0002-9440-0456 \\
Julian H\"ocker:   0000-0002-8699-3431\\
Vladimir Dyakonov:    0000-0001-8725-9573\\
Manfred Bayer:        0000-0002-0893-5949\\

\subsection{Acknowledgements}
We acknowledge the financial support by the Deutsche Forschungsgemeinschaft in the frame of the Priority Programme SPP 2196 (Project YA65/26-1) and the International Collaboration Research Center TRR160 (project B2). It was also supported by the Russian Foundation for Basic Research (Grant No. 19-52-12038 NNIO-a and 19-02-00545). M.O.N. is grateful to the Foundation for Advancement of Theoretical Physics and Mathematics "BASIS". J. H., A.B. and V.D. acknowledge financial support from the Deutsche Forschungsgemeinschaft  through the W\"urzburg-Dresden Cluster of Excellence on Complexity and Topology in Quantum Matter-ct.qmat (EXC 2147, project-id 39085490), DY18/15-1 and from the Bavarian State Ministry of Education and Culture, Science and Arts within the Collaborative Research Network "Solar Technologies go Hybrid". The work at ETH Zurich (O.N., D.N.D. and M.V.K.) was financially supported by the Swiss National Science Foundation (grant agreement 186406, funded in conjunction with SPP219 through DFG-SNSF bilateral program) and by ETH Zurich through ETH+ Project SynMatLab.

\subsection{Author contributions}
E.K., E.A.Zh., D.K., I.V.K., V.F.S., G.S.D. and N.E.K built the experimental apparatus and performed the measurements. I.V.K., V.F.S., and D.R.Y. analyzed the data. M.M.G., M.A.S., M.O.N., and  E.L.I. developed theoretical approach and performed model calculations. O.N., D.N.D., M.V.K., A.B., J.H., and V.D. grew the samples. All authors contributed to interpretation and analysis of the data. E.K., D.R.Y. and M.M.G. wrote the manuscript in close consultations with M.B., M.V.K., and V.D.

\subsection{Additional information}
Correspondence should be addressed to E.K. (email: erik.kirstein@tu-dortmund.de), D.R.Y. (email: dmitri.yakovlev@tu-dortmund.de) and M.M.G. (email: glazov@coherent.ioffe.ru).

\subsection{Competing financial interests}
Authors declare no competing financial interests.


\newpage

\clearpage
\widetext
\begin{center}
	\textbf{\large Supplementary Information:}
	
\vspace{3mm}	
	\textbf{\large The Land\'e factors of electrons and holes in lead halide perovskites: \\ universal dependence on the band gap}
	
\vspace{3mm}

E. Kirstein,$^{1}$ D. R. Yakovlev,$^{1,2}$ M. M. Glazov,$^2$ E. A. Zhukov,$^{1,2}$ D. Kudlacik,$^{1}$ I. V. Kalitukha,$^2$  V. F. Sapega,$^2$  G.~S.~Dimitriev,$^2$  M. A. Semina,$^2$ M. O. Nestoklon,$^2$   E. L. Ivchenko,$^2$ D. N. Dirin,$^3$ O. Nazarenko,$^3$  M. V. Kovalenko,$^{3,4}$ A. Baumann,$^{5}$ J.~H\"ocker,$^{5}$  V. Dyakonov,$^{5}$  and M. Bayer$^{1,2}$   

\vspace{3mm}
{\small \noindent$^1$\textit{Experimentelle Physik 2, Technische Universit{\"a}t Dortmund, 44227 Dortmund, Germany}
	
\noindent$^2$\textit{Ioffe Institute, Russian Academy of Sciences, 194021 St. Petersburg, Russia}

\noindent$^3$\textit{Department of Chemistry and Applied Biosciences,
Laboratory of Inorganic Chemistry, ETH Z\"{u}rich, 8093 Z\"{u}rich, Switzerland}

\noindent$^4$\textit{Department of Advanced Materials and
Surfaces, Laboratory for Thin Films and Photovoltaics, Empa - Swiss Federal Laboratories for Materials Science and Technology, 8600 D\"{u}bendorf, Switzerland}

\noindent$^5$\textit{Experimental Physics VI, Julius-Maximilian University of W\"urzburg, 
97074 W\"{u}rzburg, Germany}
}
\\
\end{center}


\setcounter{equation}{0}
\setcounter{figure}{0}
\setcounter{table}{0}
\setcounter{section}{0}
\setcounter{page}{1}
\renewcommand{\theequation}{S\arabic{equation}}
\renewcommand{\thefigure}{S\arabic{figure}}
\renewcommand{\bibnumfmt}[1]{[S#1]}
\renewcommand{\citenumfont}[1]{S#1}
\renewcommand{\thetable}{S\arabic{table}}
\renewcommand{\thesection}{S\arabic{section}}

\section{Theory}

In this section we present details of our theoretical approach to calculate the band structure and Zeeman effect in perovskite crystals. We outline the microscopic DFT and tight-binding methods as well as the $\bm k\cdot \bm p$-model and formulate the atomistically-inspired procedure for the estimation of the $g$-factors in perovskites.

\subsection{Preliminaries}

According to the Bloch theorem the electron wavefunctions in a crystal can be recast in the form
\begin{equation}
\label{Bloch:gen}
\Psi_{n,s;\bm k}(\bm r) = \frac{e^{\mathrm i \bm k \bm r}}{\sqrt{\mathcal V}} u_{n,s;\bm k}(\bm r),
\end{equation}
where $\bm r$ is the real space coordinate, $n$ enumerates the bands, $s$ is the spin index, $\bm k$ is the quasi-wavevector, $\mathcal V$ is the normalization volume and $u_{n,s;\bm k}(\bm r)$ is the periodic Bloch amplitude. In the presence of a weak magnetic field $\bm B$, such that all characteristic energies related to the field are much smaller than the energy separations between the bands, the Zeeman splitting of the electron states is given by the Hamiltonian with the matrix elements 
\begin{equation}
\label{Zeeman:gen}
\mathcal H_{Z,ss'} = \frac{1}{2} g_0 \mu_B (\bm \sigma_{ss'} \cdot \bm B) + \mu_B (\bm L_{ss'}\cdot \bm B).
\end{equation}
Here $s,s'=\pm 1/2$ are the spin indices, $g_0=2$ is the free-electron Land\'e factor, $\mu_B = |e|\hbar/(2m_0c)$ is the Bohr magneton with $m_0$ being free-electron mass, $e$ being the electron charge, and $c$ being the speed of light,  $\bm \sigma{=(\sigma_x,\sigma_y,\sigma_z)}$ is the vector composed of the Pauli matrices and $\bm L = -\mathrm i [\bm r\times (\partial/\partial \bm r)]$ is the angular momentum operator. 
Combining the first and second terms, Eq.~\eqref{Zeeman:gen} can be rewritten in the simple form
\begin{equation}
\label{Zeeman:gen:1SI}
\mathcal H_{Z} = \frac{\mu_B}{2} g_{\alpha\beta}  \sigma_\alpha B_\beta,
\end{equation}
where $\alpha,\beta=x,y,z$ denote the Cartesian components, $g_{\alpha\beta}$ are the elements of the $g$-factor tensor. Using the completeness relation for the Bloch functions one can recast the components of the $g$-factor tensor in the form, see e.g.,  Refs.~\cite{PhysRev.114.90_SI,birpikus_eng_SI,ivchenko2005_SI,2053-1583-2-3-034002_SI}
\begin{equation} \label{LR}
g_{\alpha\beta} { \sigma_{\alpha, ss'} } B_\beta =  \sum_{\gamma=x,y,z} B_\gamma \left(g_0 {\sigma_{ns,ns'}^{\gamma}} - \frac{2\mathrm i}{m_0} \sum_{\substack{m,t\\\alpha,\beta}} \epsilon_{\alpha\beta\gamma} \frac{p^\alpha_{ns;mt} {p^\beta_{mt;ns'}}}{E_n - E_m}\right).
\end{equation}
Here summation over the repeated subscripts is implied,  $\sigma_{ns,ns'}^{\gamma} = \langle n, s|\sigma_{\gamma}| ns'\rangle$, $\epsilon_{\alpha\beta\gamma}$ is the Levi-Civita symbol, $E_n$ and $E_m$ are the band energies at the corresponding point of the Brillouin zone, the $p^\alpha_{ns;mt}$ are the interband momentum operator matrix elements. In Eq.~\eqref{LR}  summation over all bands except for the selected band $n$ is carried out, they are enumerated by the orbital index $m$ and the spin index $t$.  For completeness we give the expression for the inverse effective mass tensor $m_{\alpha\beta}$ at a band extremum
\begin{equation}
\label{m*}
\frac{1}{m_{\alpha\beta}} = \frac{\delta_{\alpha\beta}}{m_0} + \frac{1}{m_0^2} \sum_{m,t} \frac{p^\alpha_{ns;mt} p^\beta_{mt;ns} + p^\beta_{ns;mt} p^\alpha_{mt;ns} }{E_n - E_m},
\end{equation}
where $\delta_{\alpha\beta}$ is the Kronecker symbol. Both the effective masses and Land\'e factors are determined by the set of the band structure parameters, namely, the band gaps and the interband momentum matrix elements. Thus, simultaneous calculation of both quantities by various methods allows one to improve the parametrization of the effective Hamiltonians~\cite{HW:optical_SI}. Note that if relativistic effects are included, the momentum matrix elements $p^\alpha_{ns;mt}$ should be replaced by the appropriate matrix elements of the operator $\bm \pi$ which includes the spin-orbit coupling term~\cite{birpikus_eng_SI}. 

\subsection{Band structure in the tight-binding and DFT approaches}

\begin{figure}[tb]
\includegraphics[width=0.5\linewidth]{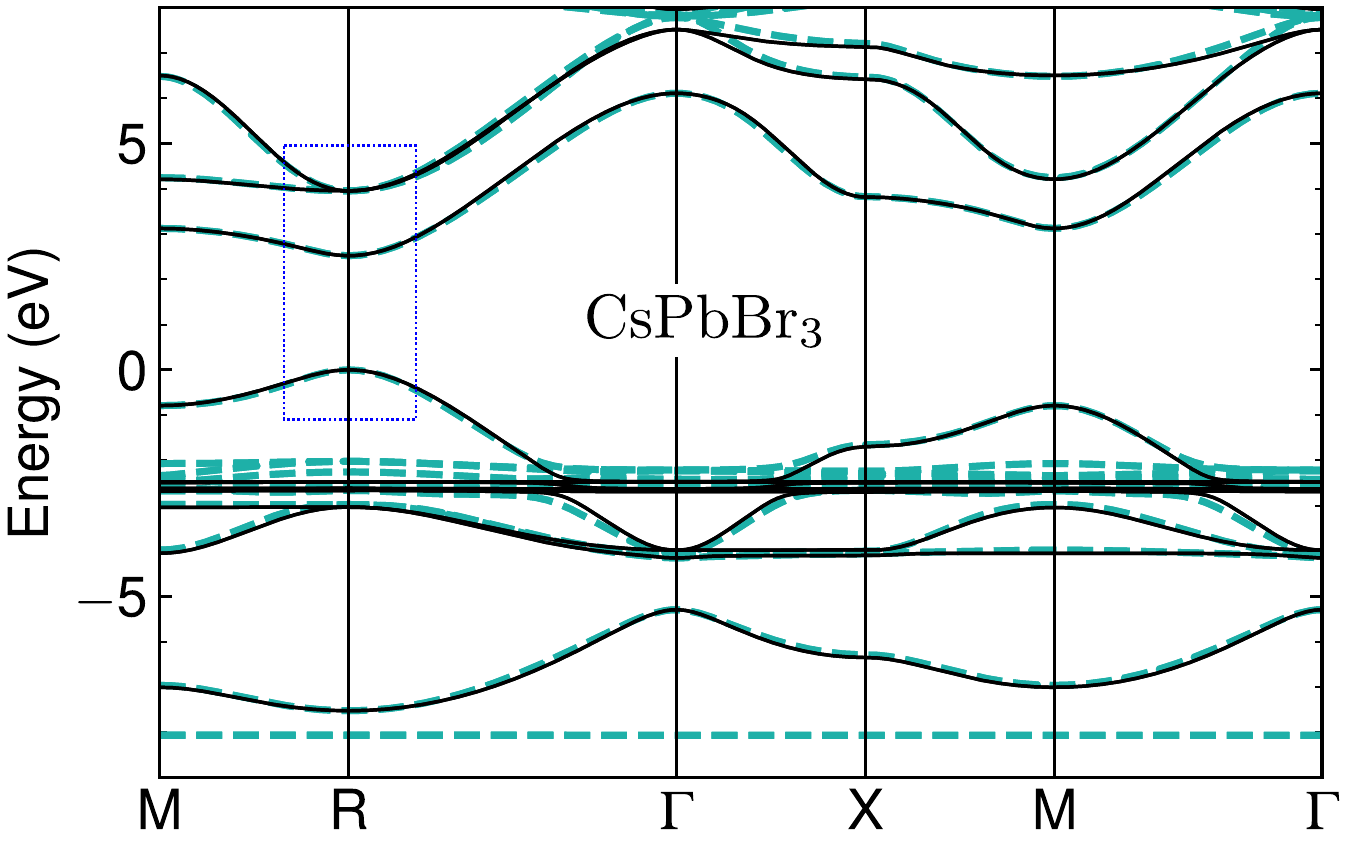}
 \caption{Comparison of the CsPbBr$_3$ band structure calculated in DFT and in the empirical tight-binding method. The DFT calculations are shown by the green dashed lines, the tight-binding results are given by the thin black lines. The blue dotted rectangle shows the topmost valence band and the lowest conduction bands in the vicinity of the $R$-point, see Fig.~\ref{fig:SI:bands:scheme}(a). }\label{fig:TBvsExpt}
\end{figure}

For a realistic description of the band structure of perovskites we use a tight-binding model based on DFT calculations, see Ref.~\cite{Nestoklon:2021wg_SI}. Such a combination of  \emph{ab initio} and \emph{empirical} methods allows us to gain access to the details of the band structure and provides ground for a simplified $\bm k\cdot \bm p$-modeling of the bands. In our analysis, we consider the high-symmetry cubic structure. This is a good starting point: the energy bands of lower symmetry crystal phases may be then described as a folded and distorted band structure of this most symmetric phase \cite{even_pedestrian_2015_SI}. The standard approach for the description of the electronic properties of the cubic phase of the prototype hybrid organic-inorganic perovskite CH\textsubscript{3}NH\textsubscript{3}PbI\textsubscript{3} is to consider its all-inorganic analogue, CsPbI\textsubscript{3} \cite{even_pedestrian_2015_SI}. The DFT calculations are performed using the WIEN2k package \cite{wien2k_SI} with the modified Becke-Johnson exchange-correlation potential \cite{Tran09_SI} in Jishi parametrization \cite{Jishi14_SI}, for details see Ref.~\cite{Nestoklon:2021wg_SI}. Note that for cubic CsPbI\textsubscript{3} with the lattice constant {$6.289$}~\AA, this approach gives the band gap $E_g=1.366$~eV, while the experimental value is $E_g=1.65$~eV \cite{Yuan20_SI}. The difference between the DFT results and experimental data for this material is attributed to the renormalization of the band structure by the electron-phonon interaction \cite{Cardona01_SI} which is estimated as hundreds of meV \cite{Wiktor17_SI}. The DFT calculations of the band structure of cubic CsPbBr\textsubscript{3} and CsPbCl\textsubscript{3} are performed in the same way. The lattice constants, $5.992$~\AA\ and $5.605$~\AA, respectively, are taken from Ref.~[\citenum{Jishi14_SI}]. The value of the radius of the muffin tin used is $R_{\rm{MT}}=2.5$~Bohr, except for Cl in which case $R_{\rm{MT}}=2.37$~Bohr. The parameters controlling the numerical precision are $R_{\rm{MT}}\cdot K_{\rm{MAX}}=13$ and $E_{\rm{MAX}}=14$. With these parameters, the band gap for the cubic phase is $E_g=2.52$~eV for CsPbBr\textsubscript{3} and $E_g=3.09$~eV for CsPbCl\textsubscript{3}. The dispersions for the CsPbBr$_3$ cubic crystal calculated by the DFT approach are shown by the green dashed curves in Fig.~\ref{fig:TBvsExpt}.

For the tight-binding calculations, we use the empirical tight-binding model with the $sp^3d^5s^*$ basis in the nearest neighbor approximation. It gives a precise description of the band structure of bulk III--V \cite{Jancu98_SI} and group IV \cite{Niquet09_SI} semiconductors. Recently, it has been shown that the extended $sp^3d^5s^*$ tight-binding method can be used to describe the band structure of inorganic perovskites with a meV-range precision \cite{Nestoklon:2021wg_SI}. For CsPbI\textsubscript{3} we refer to the parameters from Ref.~\cite{Nestoklon:2021wg_SI}. For the two other perovskites, CsPbBr\textsubscript{3} and CsPbCl\textsubscript{3}, we use the results of the tight-binding fit to the DFT approach outlined in Ref.~\cite{Nestoklon:2021wg_SI}, they are given in Table~\ref{tbl:TB_par}; these materials are not stable in the cubic phase so that corresponding experimental data are not available. Note that the relative error in the DFT band gap is significantly smaller for Br- and Cl-based materials since their band gap is much larger. The corresponding tight-binding dispersion curves for CsPbBr$_3$  are shown in Fig.~\ref{fig:TBvsExpt} by the solid thin black lines and demonstrate good agreement with the DFT calculations.

\begin{table}\caption{Tight-binding parameters fitted to the DFT calculations. All values are given in eV. In addition to the parameters presented in the table, the parameters $s_cs^*_a\sigma$, $s^*_ad_c\sigma$, $p_cd_a\sigma$, $p_ad_c\pi$ and those involving the $s_a$ and $s^*_c$ orbitals are taken to be zero. The parameters of CsPbI\textsubscript{3} are reproduced from Ref. \cite{Nestoklon:2021wg_SI}.}

\label{tbl:TB_par}
\begin{tabular*}{0.4\linewidth}{@{\extracolsep{\fill}}lrrr}
 \hline
 \hline
 &CsPbI\textsubscript{3} & CsPbBr\textsubscript{3} & CsPbCl\textsubscript{3} \\
 \hline
 $         E_{sc}$ & $   -5.7767$ & $   -4.9645$ & $   -4.5816$\\
 $       E_{s^*a}$ & $   19.6780$ & $   19.7944$ & $    8.6310$\\
 $         E_{pa}$ & $   -2.3350$ & $   -2.5653$ & $   -3.4496$\\
 $         E_{pc}$ & $    4.4825$ & $    5.5804$ & $    6.4591$\\
 $         E_{da}$ & $   10.8491$ & $   12.9468$ & $   14.1585$\\
 $         E_{dc}$ & $   13.9357$ & $   15.9568$ & $   15.9638$\\
 $   s_cp_a\sigma$ & $    1.0421$ & $    1.0288$ & $    1.1429$\\
 $ s^*_ap_c\sigma$ & $    2.7092$ & $    2.5482$ & $    1.3349$\\
 $   s_cd_a\sigma$ & $    0.3749$ & $   -0.7094$ & $    0.7972$\\
 $       pp\sigma$ & $   -1.8838$ & $   -1.8488$ & $   -2.0710$\\
 $          pp\pi$ & $    0.1955$ & $    0.1990$ & $    0.1183$\\
 $   p_ad_c\sigma$ & $    1.0341$ & $   -1.1460$ & $    1.4326$\\
 $      p_cd_a\pi$ & $   -0.7960$ & $   -0.7768$ & $   -0.9618$\\
 $       dd\sigma$ & $   -1.1231$ & $   -1.1429$ & $   -1.1768$\\
 $          dd\pi$ & $    2.0000$ & $    2.0000$ & $    2.0000$\\
 $       dd\delta$ & $   -1.4000$ & $   -1.4000$ & $   -1.4000$\\
 $     \Delta_a/3$ & $    0.3250$ & $    0.0944$ & $    0.2290$\\
 $     \Delta_c/3$ & $    0.4892$ & $    0.5469$ & $    0.7071$\\
\hline
\hline
\end{tabular*}
\end{table}

\subsection{Effective Hamiltonian model}\label{sec:effective}

\begin{figure}
\includegraphics[width=0.7\textwidth]{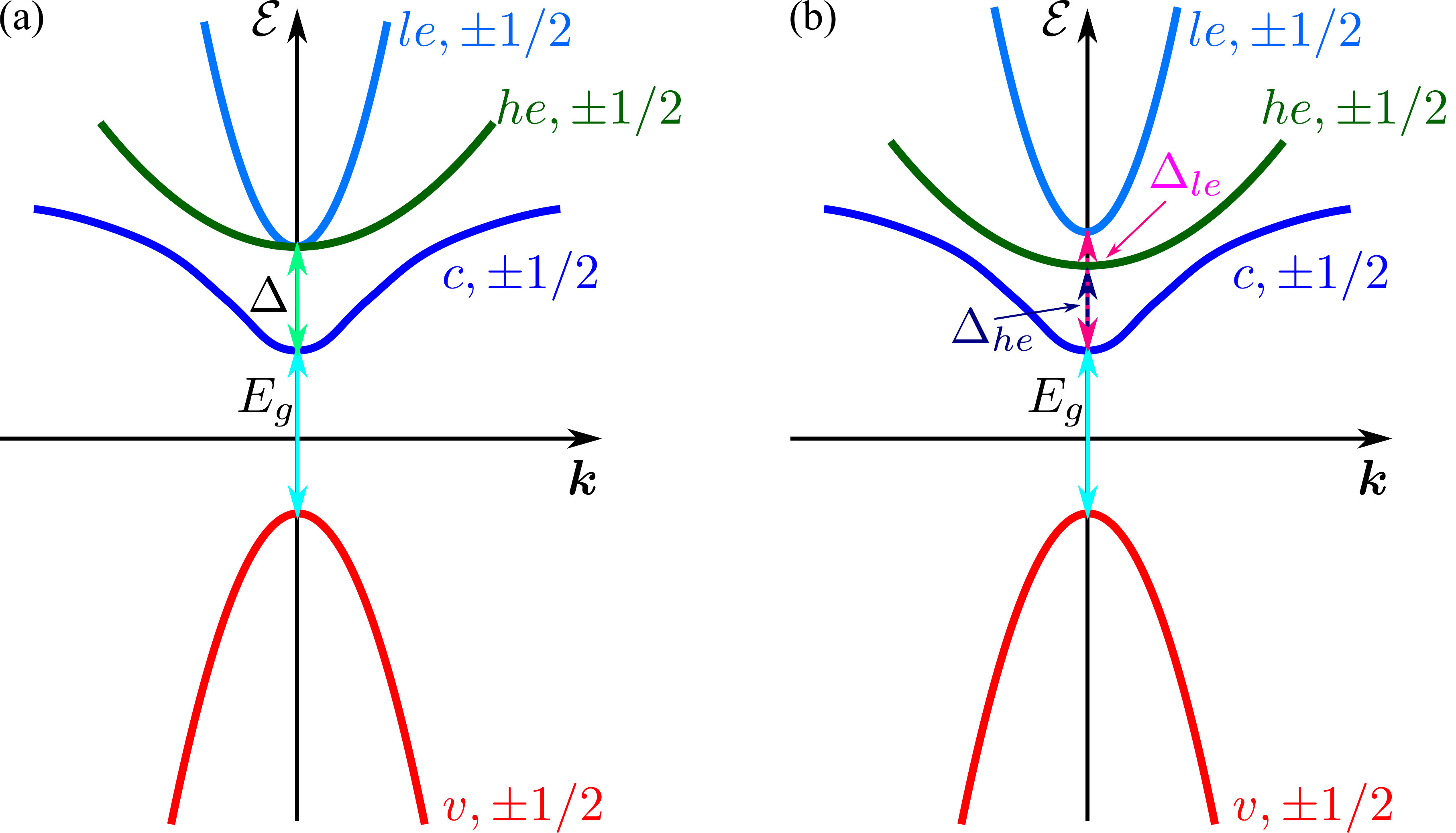}
\caption{Schematic illustration of the band structure of a bulk (a) cubic perovskite crystal in the vicinity of the R-point of the Brillouin zone and (b) tetragonal perovskite crystal in the vicinity of the $\Gamma$-point.}\label{fig:SI:bands:scheme}
\end{figure}

In this section we present the basic information about the $\bm k\cdot \bm p$ model applied to calculation of the conduction and valence band states in bulk perovskites and their key properties: the effective masses and $g$-factors. Following Refs.~\cite{yu2016_SI,sercel_exciton_2019_SI,sercel_quasicubic_2019_SI} we start our description with the minimum model which includes the topmost two-fold degenerate valence band and the nearest conduction bands. In the high-temperature cubic crystalline modification of the perovskites the direct band gap is formed at the R-point of the Brillouin zone (corner of the cube in the $[111]$ direction). The schematics of the band structure in the vicinity of the R-point are shown in Fig.~\ref{fig:SI:bands:scheme}(a). For a tetragonal perovskite the bands are folded and the band gap is located at the $\Gamma$ point of the Brillouin zone due to the band folding~\cite{PhysRevB.86.205301_SI}, Fig.~\ref{fig:SI:bands:scheme}(b).

In both the cubic and tetragonal cases the valence band Bloch amplitudes can be recast in the form 
\begin{subequations}
\label{Bloch:amp}
\begin{equation}
\label{Bloch:v}
\mbox{valence band:}\quad \begin{cases}
u_{v,\frac12}(\bm r) = {\rm i} \mathcal S(\bm r) \hspace{-1 mm}\uparrow,\\
u_{v,-\frac12}(\bm r) = {\rm i} \mathcal S(\bm r) \hspace{-1 mm}\downarrow,\\
\end{cases}
\end{equation}
where $\uparrow, \downarrow$ denote the basic spinors, and $\mathcal S(\bm r)$ is the invariant function. The conduction band states are formed from the three orbital Bloch amplitudes $\mathcal X(\bm r)$, $\mathcal Y(\bm r)$, and $\mathcal Z(\bm r)$ which transform as the corresponding coordinates: in the cubic modification the axes $x\parallel [100]$, $y\parallel [010]$, $z\parallel [001]$ are equivalent, and in the tetragonal perovskite the $z$-axis is the $C_4$-axis with $x$ and $y$ being equivalent. The Bloch states are determined by the interplay of the crystalline splitting and spin-orbit interaction and can be described as
\begin{equation}
\label{Bloch:vi}
\mbox{bottom conduction band (c.b.):}\quad \begin{cases}
u_{c,\frac12}(\bm r) = -\sin{\vartheta} {\cal Z}(\bm r)\hspace{-1 mm}\uparrow - \cos{\vartheta} \cfrac{\X(\bm r) + \mathrm i \Y(\bm r)}{\sqrt{2}}\hspace{-1 mm}\downarrow,\\
u_{c,-\frac12}(\bm r) =\sin{\vartheta} \Z(\bm r)\hspace{-1 mm}\downarrow - \cos{\vartheta} \cfrac{\X(\bm r) - \mathrm i \Y(\bm r)}{\sqrt{2}}\hspace{-1 mm}\uparrow,\\
\end{cases}
\end{equation}
\begin{equation}
\label{Bloch:vii}
\mbox{excited (light electron) c.b.:}\quad \begin{cases}
u_{le,\frac12}(\bm r) = \cos{\vartheta} \Z(\bm r)\hspace{-1 mm}\uparrow - \sin{\vartheta} \cfrac{\X(\bm r) + \mathrm i \Y(\bm r)}{\sqrt{2}}\hspace{-1 mm}\downarrow,\\
u_{le,-\frac12}(\bm r) =\cos{\vartheta} \Z(\bm r)\hspace{-1 mm}\downarrow + \sin{\vartheta} \cfrac{\X(\bm r) - \mathrm i \Y(\bm r)}{\sqrt{2}}\hspace{-1 mm}\uparrow,\\
\end{cases}
\end{equation}
\begin{equation}
\label{Bloch:viii}
\mbox{excited (heavy electron) c.b.:}\quad \begin{cases}
u_{he, \frac32}(\bm r) = -\cfrac{\X(\bm r) + \mathrm i \Y(\bm r)}{\sqrt{2}}\hspace{-1 mm}\uparrow,\\
u_{he,- \frac32}(\bm r) = \cfrac{\X(\bm r) - \mathrm i \Y(\bm r)}{\sqrt{2}}\hspace{-1 mm}\downarrow,\\
\end{cases}
\end{equation}
\end{subequations} 

In the cubic modification the light and heavy electron states at the $R$-point are degenerate, Fig.~\ref{fig:SI:bands:scheme}(a). In the tetragonal modification a splitting between the light and heavy electrons arises, as Fig.~\ref{fig:SI:bands:scheme}(b).  We introduce the band gap $E_g$ as the energy gap between the bottom conduction band and the valence band, $E_g+\Delta_{le}$ is the gap between the light electron band and valence band, $E_g+\Delta_{he}$ is the gap between the heavy electron band and the valence band, see Fig.~\ref{fig:SI:bands:scheme}(b). The interband momentum matrix elements defined by (we assume that $p_\parallel$ and $p_\perp$ are real) are given by
\begin{equation} \label{pcv}
p_\perp = {\rm i} \langle \X |p_x|\S\rangle = {\rm i}\langle \Y |p_y|\S\rangle, \\ \quad p_\parallel = {\rm i} \langle \Z |p_z|\S\rangle . 
\end{equation}
The parameter $\vartheta$  determines the relation between the crystalline splitting and the spin-orbit interaction. Naturally, in the cubic approximation $\cos{\vartheta} = \sqrt{2/3}$, $\sin{\vartheta}=1/\sqrt{3}$, $\Delta_{le} = \Delta_{he}\equiv \Delta$, and $p_\parallel = p_\perp \equiv p$. In the quasi-cubic approximation~\cite{birpikus_eng_SI,sercel_exciton_2019_SI,sercel_quasicubic_2019_SI} the anisotropy parameters can be expressed in terms of the bare spin-orbit splitting energy $\Delta_{so}$ and the crystal field splitting $\Delta_c$ as
\begin{equation}
\label{aniso:param}
\Delta_{le} = \sqrt{\Delta_{so}^2+ \Delta_c^2-\frac{2}{3} \Delta_{so}\Delta_c}, \quad \Delta_{he} = \frac{\Delta_{le}+ \Delta_{so}+\Delta_c}{2}, \quad \tan{2\vartheta} = \frac{2\sqrt{2} \Delta_{so}}{\Delta_{so} - 3\Delta_c}.
\end{equation}

Within this minimum model we calculate the effective masses and $g$-factors for electrons and holes. For the electron at the bottom of the conduction band ($c,\pm 1/2$), see Fig.~\ref{fig:SI:bands:scheme}, we obtain:  
\begin{subequations}
\label{cb}
\begin{equation}
\label{cb:par}
\frac{1}{m_{e\parallel}}= \frac{1}{m_0} + \frac{2 p_\parallel^2 }{m_0^2} \frac{\sin^2\vartheta}{E_g},
\end{equation}
\begin{equation}
\label{cb:perp}
\frac{1}{m_{e\perp}}= \frac{1}{m_0} + \frac{p_\perp^2 }{m_0^2} \frac{\cos^2\vartheta}{E_g}.
\end{equation}
\end{subequations}
Here the symbols $\parallel$ and $\perp$ denote the direction of the electron propagation, namely, along and perpendicular to the $C_4\parallel z$ axis, respectively. In the valence band the hole effective masses take the form
\begin{subequations}
\label{vb}
\begin{equation}
\label{vb:par}
-\frac{1}{m_{h\parallel}} = \frac{1}{m_{v.b.\parallel}}= \frac{1}{m_0} - \frac{2p_\parallel^2}{m_0^2}  \left(\frac{\sin^2\vartheta}{E_g} +\frac{\cos^2\vartheta}{E_g+\Delta_{le}} \right),
\end{equation}
\begin{equation}
\label{vb:perp}
-\frac{1}{m_{h\perp}} = \frac{1}{m_{v.b.\perp}}= \frac{1}{m_0} - \frac{p_\perp^2}{m_0^2}  \left(\frac{\cos^2\vartheta}{E_g} +\frac{\sin^2\vartheta}{E_g+\Delta_{le}} +\frac{1}{E_g+\Delta_{he}} \right).
\end{equation}
\end{subequations}
Note that the hole masses, $m_h$, have opposite signs as compared to the valence band electron masses, $m_{v.b.}$. Evidently, for the cubic crystal the effective masses and Land\'e factors become isotropic. In particular,
\begin{equation}
\label{masses:iso}
\frac{1}{m_{e\parallel}} = \frac{1}{m_{e\perp}} = \frac{1}{m_0} + \frac{2p^2}{m_0^2 E_g}, \quad -\frac{1}{m_{h\parallel}} =- \frac{1}{m_{h\perp}} = \frac{1}{m_0} - \frac{2p^2}{3m_0^2}\frac{3E_g + \Delta}{E_g(E_g+\Delta)}.
\end{equation} 
Equations~\eqref{vb} transform into the standard expression (2.48) of Ref.~\cite{ivchenko2005} for cubic III-V semiconductors when neglecting the non-parabolicity and making the natural replacements $E_g \to - (E_g+\Delta)$, $E_g+\Delta \to - E_g$ that are related to the difference in the band order in perovskites and GaAs-like crystals.

We now turn to the Land\'e factors. For the bottom conduction band one has
\begin{subequations}
\label{g:cb}
\begin{equation}
\label{g:cb:par}
\hspace{-16 mm}g_{e\parallel} = - \frac23+ \frac{2p_\perp^2}{m_0}  \frac{\cos^2\vartheta}{E_g},
\end{equation}
\begin{equation}
\label{g:cb:perp}
g_{e\perp} = - \frac23+ \frac{2\sqrt{2} p_\parallel p_\perp}{m_0} \frac{\cos{\vartheta} \sin{\vartheta} }{E_g}.
\end{equation}
\end{subequations}
Here the subscripts $\parallel$ and $\perp$ of the Land\'e factors denote the direction of the magnetic field with respect to the $C_4$ axis. In these expressions we took into account the difference from the value of $2$ of the free-electron spin contribution to the $g$-factor due to the mixed form of wavefunctions~\eqref{Bloch:vi}, see the first term on the r.h.s. of Eq.~\eqref{LR}. For the valence band we obtain
\begin{subequations}
\label{g:vb}
\begin{equation}
\label{g:vb:par}
g_{h\parallel} = 2- \frac{2p_\perp^2}{m_0}  \left(\frac{\cos^2\vartheta}{E_g} +\frac{\sin^2\vartheta}{E_g+\Delta_{le}} - \frac{1}{E_g + \Delta_{he}}\right),
\end{equation}
\begin{equation}
\label{g:vb:perp}
g_{h\perp} = 2- \frac{2\sqrt{2} p_\parallel p_\perp}{m_0} \cos{\vartheta} \sin{\vartheta} \left(\frac{1}{E_g} - \frac{1}{E_g+\Delta_{le}}  \right).
\end{equation}
\end{subequations}
Note that the valence band $g$-factors in the electron and hole representation have the same sign because the transformation from the electron to the hole representation includes both a change in the sign of energy and the time reversal. We define the Land\'e factor in such a way that, e.g., for $\bm B \parallel z$ the splitting $E_{+1/2} - E_{-1/2}$ between the states with spin projection $+1/2$ and $-1/2$ onto the $z$ axis is given by $g_{e\parallel} \mu_B B_z, g_{h\parallel} \mu_B B_z$, see Eq.~\eqref{Zeeman:gen:1SI}.

The expressions for the valence band $g$-factor in the cubic limit transform into the well-known formula (2.48) of Ref.~\cite{ivchenko2005_SI} with the same replacements $E_g \to - (E_g+\Delta)$, $E_g+\Delta \to - E_g$  as for the effective mass. The contributions due to the $\bm k\cdot\bm p$ interaction of the conduction and valence bands are also in agreement with  Yu~\cite{yu2016_SI}, both in terms of magnitudes and signs. When comparing with Ref.~\cite{yu2016_SI}, one has to keep in mind that in the notations of Yu the energy is reckoned from the heavy electron band and $P_{\parallel,\perp} = (\hbar/m_0)p_{\parallel,\perp}$. Also $\hbar^{-2}$ is omitted in Eqs.~(11)-(14) of Ref.~\cite{yu2016_SI}. 

Importantly, in Ref.~\cite{yu2016_SI} the remote band contributions to the conduction band  Land\'e factors were included through the `magnetic' Luttinger parameters $\kappa_{1,2}$. The importance of the remote band contributions for the conduction band parameters is highlighted by our microscopic calculations, see below. To illustrate their role let us estimate, within the cubic approximation, the contribution of the remote band with the orbital Bloch functions $\mathcal X\mathcal Y$, $\mathcal X\mathcal Z$, and $\mathcal Y\mathcal Z$ ($F_1^+$ or $R_5^+$) to the effective mass and $g$-factor of the conduction band electron. The importance of this band will be clarified below from the comparison of the $\bm k\cdot \bm p$-method with the atomistic approaches. Due to their even parity, these bands do not contribute to the hole mass and Land\'e factor. We denote
\begin{equation}
\label{q:me}
q = \langle \mathcal X\mathcal Y|p_x |\mathcal Y\rangle = \langle \mathcal X\mathcal Y|p_y |\mathcal X\rangle,~\mbox{etc.},
\end{equation}
and select the phases of the wavefunctions in such a way that $q$ is real. We also neglect the spin-orbit splitting of the remote band as compared to the distance $E_g'$ between the remote band and the conduction band.
The contribution to the inverse effective mass reads 
\begin{equation}
\label{dm:remote}
\Delta \left(m_e^{-1}\right) = \frac{4}{3} \frac{q^2}{m_0^2 E_g'} \quad\Rightarrow \quad \frac{1}{m_{e}}= \frac{1}{m_0} + \frac{2 p^2 }{m_0^2} \frac{1}{3E_g} + \frac{4}{3} \frac{q^2}{m_0^2 E_g'},
\end{equation}
while the contribution to the Land\'e factor takes the form
\begin{equation}
\label{dg:remote}
\Delta g_e = - \frac{4}{3} \frac{q^2}{m_0 E_g'}.
\end{equation}
In the cubic approximation this correction is isotropic, the difference of the remote bands contribution to the Land\'e factor $\Delta g_{e\parallel} - \Delta g_{e\perp}$ appears due to the crystalline splitting of the remote $F_1^+$ orbitals and the anisotropy of the matrix elements in Eq.~\eqref{q:me}.

\subsection{Results}

To compute the $g$-factor in the empirical tight-binding (ETB) model we directly use Eq. \eqref{LR}. We construct the tight-binding Hamiltonian in the $R$-point of the Brillouin zone, and find the eigenvalues as well as the eigenvectors that represent the energies and wave functions at the $R$-point. The resulting eigenvectors should be ``symmetrized'' to form the canonical basis of irreducible representations in the $R$ point ({$R_6^+$} for the valence band and {$R_6^-$} for the conduction band \cite{even_pedestrian_2015_SI}), in order to exclude an arbitrary phase of the computed eigenvectors from the $g$-factor values. Then, in this basis, we calculate the matrix elements of the spin operator and the velocity operator in the standard manner \cite{Graf95_SI}, and use them in Eq.~\eqref{LR} to find the $g$-factors. 

\begin{table}[h]\caption{Band gap, spin-orbit splitting, effective masses and $g$-factors for different perovskite materials extracted from DFT and ETB calculations. The band gap and spin-orbit splitting from DFT are reproduced in the ETB exactly, while there is a small difference in the masses. The $g$-factors are extracted only from the ETB calculations. Subscripts $e$ and $h$ denote the bottom conduction band electron states and the top valence band hole states.}
\label{tbl:g_ETB}
\begin{tabular*}{\linewidth}{@{\extracolsep{\fill}}ccccccccc}
 \hline
 \hline
       & $E_g$ (eV)    &$\Delta$  (eV) & $m_h/m_0$ (DFT) 
        & $m_h/m_0$ (ETB) 
                                      & $m_e/m_0$ (DFT) 
                                               
                                                          & $m_e/m_0$ (ETB) 
                                                                    &$g_h$ (ETB) 
                                                                              & $g_e$ (ETB) \\
 \hline
 CsPbI\textsubscript{3} 
   & $ 1.366$ & $1.266$ & $  0.16$& $0.191$ &$ 0.18$ &  $0.184$ & $-0.108$ & $1.660$\\
 CsPbBr\textsubscript{3}
   & $ 2.520$ & $1.431$ & $  0.26$& $0.298$ &$ 0.30$ &  $0.291$ & $1.343$ & $0.127$\\
 CsPbCl\textsubscript{3}
   & $ 3.090$ & $1.526$ & $  0.28$& $0.348$ &$ 0.36$ &  $0.398$ & $1.527$ & $-0.080$\\
\hline
\hline
\end{tabular*}
\end{table}

\begin{figure}[h]
\includegraphics[width=0.75\textwidth]{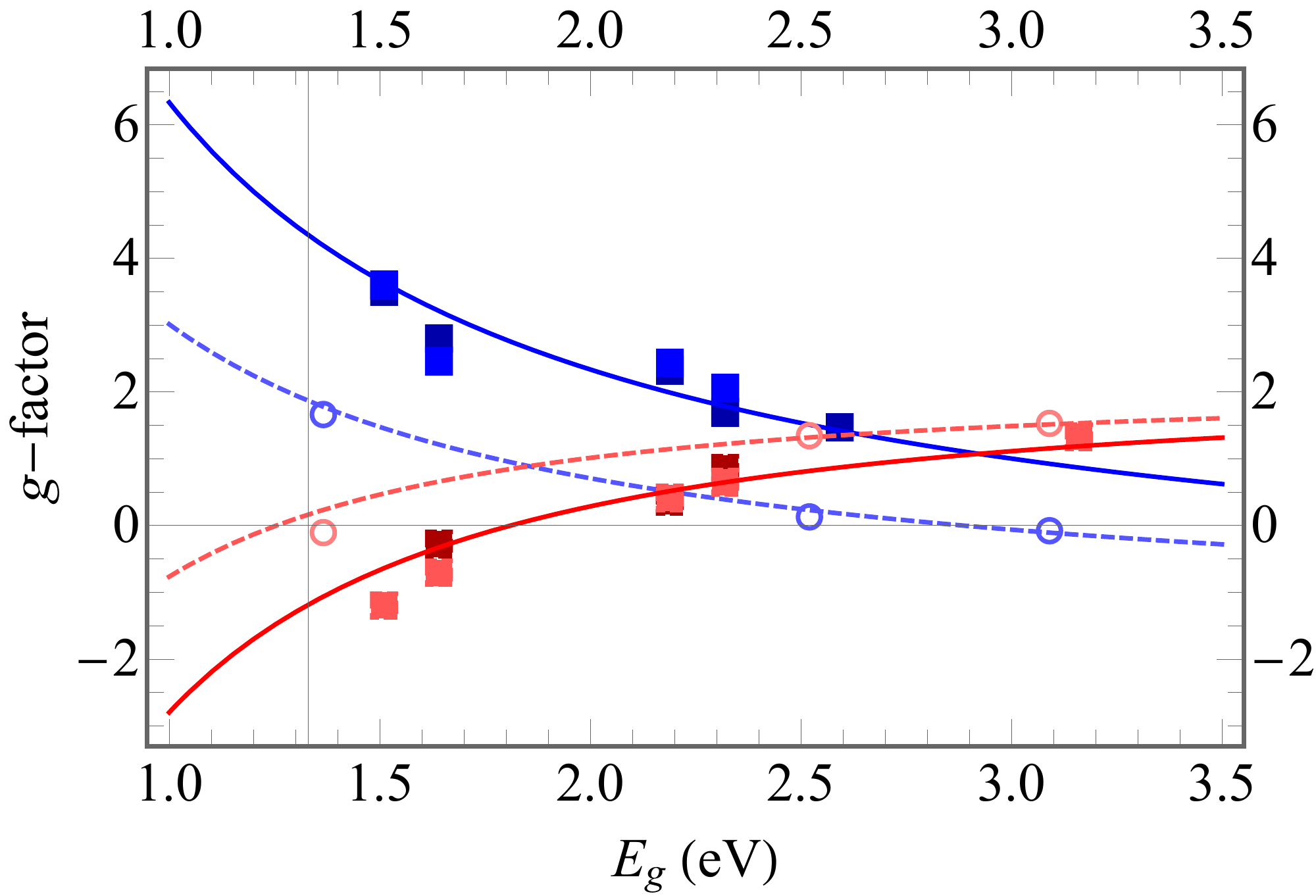}
\caption{Summary of the calculated and measured Land\'e factors for electrons and holes in the perovskites as function of the band gap. Blue symbols and lines show the electron $g$-factors, red symbols and lines show the hole $g$-factors. Open circles are ETB calculations from Table~\ref{tbl:g_ETB}. Dashed lines are calculated within the atomically-inspired $\bm k\cdot\bm p$-approach after Eqs.~\eqref{gv:cub} and \eqref{gc:cub} with $\Delta = 1.5$~eV, $P = \hbar p/m_0 = 5.1$~eV\AA, and $\Delta g_e=-0.94$, closely matching the ETB calculations. Filled squares show the experimental data (see main text for details).
Solid lines are calculated within the $\bm k\cdot\bm p$-approach after Eqs.~\eqref{gv:cub} and \eqref{gc:cub} with $\Delta = 1.5$~eV, $P = \hbar p/m_0 = 6.8$~eV\AA, and $\Delta g_e=-1$ in reasonable agreement with the experiment. }
\label{fig:SI:G-summary}
\end{figure}

The calculated values of the Land\'e factors are summarized in Tab.~\ref{tbl:g_ETB}. It also presents the values of the effective masses found by fitting the numerically calculated dispersion curves with parabolas in the vicinity of the $R$-point. Note that the direct application of Eq.~\eqref{m*} yields similar values for the conduction band effective masses, but larger values for the valence band effective masses because of the $k^2$ diagonal terms appearing in the tight-binding Hamiltonian, see the discussion in Ref.~\cite{Graf95_SI}. We note that the $g$-factor dependence on the band gap corresponds well to the experiment. The incomplete agreement with the measured data of the $g$-factors are mainly related to the difference in the band gap energies of our DFT and ETB atomistic calculations and those observed for the studied crystals, as well as to limitations in the extraction of the momentum matrix elements in the state-of-the-art DFT$\to$ ETB procedures. The open circles in Fig.~\ref{fig:SI:G-summary} show the Land\'e factors from the ETB procedure as function of the band gap energy. It is seen that the atomistic approach gives reasonable values of the Land\'e factor with the right trends: the electron $g$-factor decreases with increasing band gap energy $E_g$, while the hole $g$-factor increases with increasing $E_g$.

\begin{figure}[hbt]
\includegraphics[width=\textwidth]{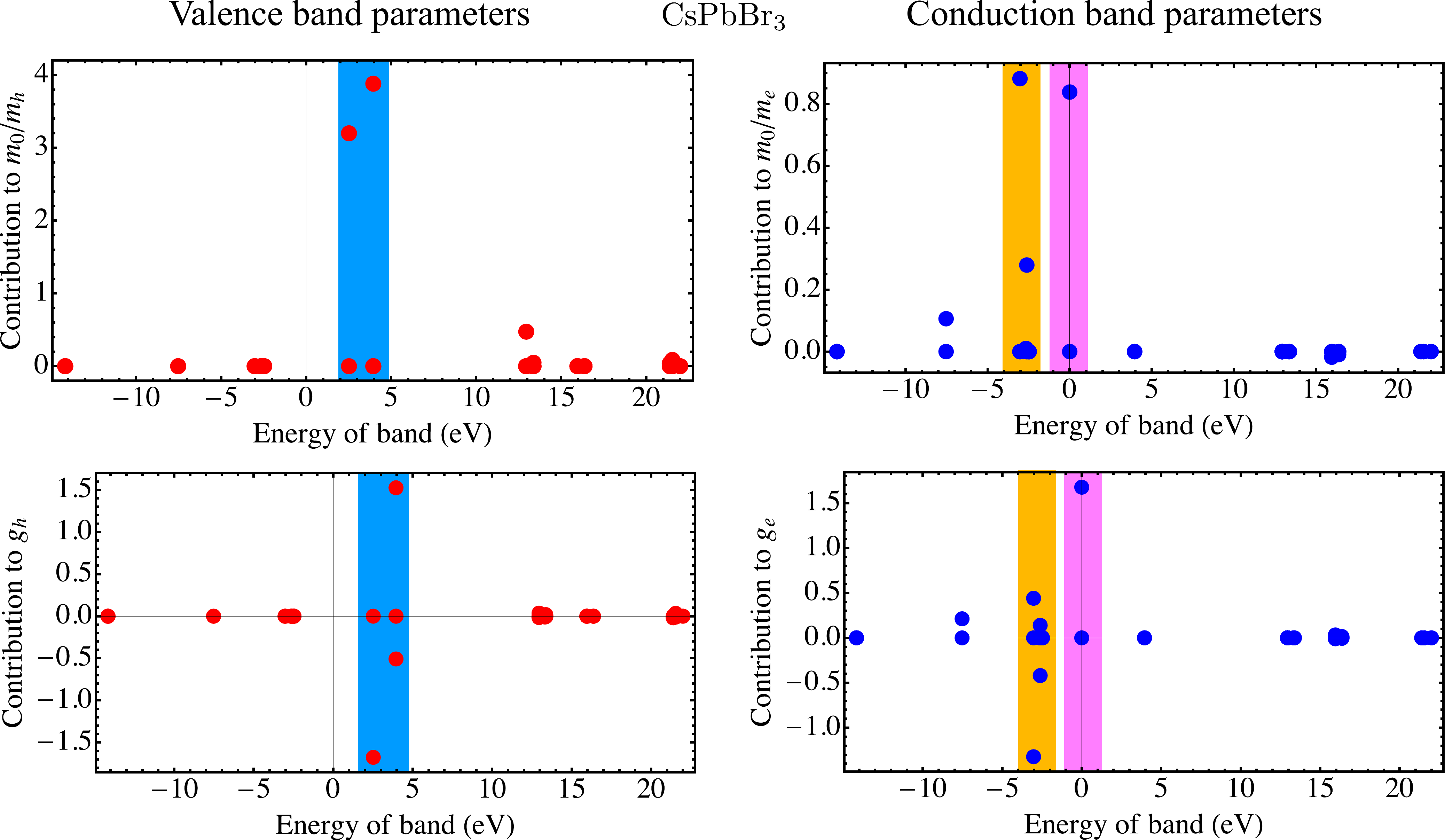}
\caption{Contributions to the effective masses and $g$-factors from various bands calculated  for CsPbBr$_3$ after Eqs.~\eqref{LR} and \eqref{m*}. Blue, magenta and yellow rectangles show the energy ranges of the conduction band, the topmost valence band and a bunch of remote valence bands of $F_1^+$ symmetry, respectively. The contributions are given in arbitrary units.}
\label{fig:SI:contributions}
\end{figure}

To gain further insight into the key band parameters and to provide atomistic empirical expressions for the $g$-factors within the effective Hamiltonian model we have analyzed the contributions of the different bands to the effective masses and $g$-factors, Fig.~\ref{fig:SI:contributions}. Namely, we evaluated the different terms in Eqs.~\eqref{LR} and \eqref{m*} resulting from band mixing and plotted them in arbitrary units as function of the energy $E_m$ of the corresponding state involved in the summation. Larger values indicate larger contributions to the inverse effective mass and Land\'e factor. The calculations clearly show that, for the valence band, the effective mass and the $g$-factor are dominated by the contribution of the conduction band (blue rectangle in Fig.~\ref{fig:SI:contributions}, left panels). For the bottom conduction band, both the topmost valence band (magenta) and the remote valence bands (yellow) are important. Furthermore, it follows from our parametrization that the Kane matrix element $P=(\hbar/m_0)p$ depends weakly on the material ranging from $4.4$~eV$\cdot$\AA~to $5.5$~eV$\cdot$\AA. This enables us to use the atomically-inspired effective $\bm k\cdot \bm p$-approach formulated in Sec.~\ref{sec:effective} to evaluate the Land\'e factors for the perovskite crystals in the cubic phase. To that end we use Eq.~\eqref{g:vb}
\begin{equation}
\label{gv:cub}
g_{h} = 2- \frac{4p^2}{3m_0}  \left(\frac{1}{E_g} -\frac{1}{E_g+\Delta}\right),
\end{equation}
for the hole $g$-factor and a combination of Eqs.~\eqref{g:cb} and \eqref{dg:remote}
\begin{equation}
\label{gc:cub}
g_{e} = -\frac{2}{3}+ \frac{4p^2}{3m_0E_g}  + \Delta g_e,
\end{equation}
for the electron $g$-factor. We recall that $p=p_\parallel = p_\perp$ in the cubic case. Taking $\Delta = 1.5$~eV, $P=\hbar p/m_0 =5.1$~{eV$\cdot$\AA}, and $\Delta g_e= -0.94$ we simultaneously reproduce the band gap dependence of the Land\'e factors found in the ETB approach, see the dashed lines in Fig.~\ref{fig:SI:G-summary}. Equations~\eqref{gv:cub} and \eqref{gc:cub} clearly highlight the physics behind the dependences: The contributions to the $g$-factors due to the conduction band-valence band mixing have opposite signs for the electron (positive) and hole (negative) and decrease in absolute value with increasing band gap energy. That is why the electron $g$-factor decreases with increasing  $E_g$ (from large positive values at small $E_g$ to $-2/3+\Delta g_e$) and the hole $g$-factor increases (from large negative values, it passes through zero and reaches $+2$ at large band gaps).

Choosing reasonable values $P=\hbar p/m_0 = 6.8$~{eV$\cdot$\AA, $\Delta = 1.5$}~eV, and $\Delta g_e = -1$, we obtain good agreement with the experimental data, see the solid lines in Fig.~\ref{fig:SI:G-summary}.

Using Eqs.~\eqref{g:cb} and \eqref{g:vb} we evaluate also the bright exciton $g$-factor which describes the splitting of the exciton radiative doublet into circularly polarized components, by $g_X=g_e+g_h$. For the magnetic field along the main axis we have
\begin{subequations}
\label{g:x}
\begin{align}
g_{X\parallel} = {\frac43} - \frac{2p_\perp^2}{m_0}  \left(\frac{\sin^2\vartheta}{E_g+\Delta_{le}} - \frac{1}{E_g + \Delta_{he}}\right) + \Delta g_{e},\\
g_{X\perp} = {\frac43} + \frac{2\sqrt{2} p_\parallel p_\perp}{m_0}  \frac{\cos{\vartheta} \sin{\vartheta}}{E_g+\Delta_{le}}  + \Delta g_{e}.
\end{align}
\end{subequations}
It is noteworthy that the contributions to the individual $g$-factors due to the $\bm k\cdot\bm p$-mixing of the valence band with the bottom conduction band $\propto 1/E_g$  cancel in the exciton $g$-factor.

\subsection{$g$-factor anisotropy in CsPbBr$_3$}

It is instructive to analyze in more detail the anisotropy of the $g$-factor components for the case of CsPbBr$_3$, where this anisotropy is most clearly pronounced. in particular, the experiment shows that
\begin{equation}
\label{g:exper:anis}
g_{e\parallel} = 1.69, \quad g_{e\perp} = 2.06 ;  \quad g_{h\parallel}=0.85, \quad g_{h\perp}=0.65.
\end{equation}
Our model has the following parameters: $p_\parallel$, $p_\perp$, $\Delta_{le}$, $\Delta_{he}$, and $\vartheta$. The interband momentum matrix elements $p_\parallel$ and $p_\perp$ can be considered as independent parameters, but the three remaining parameters which deteremine the conduction band structure, can be expressed by virtue of Eq.~\eqref{aniso:param} via two energies, namely the spin-orbit energy $\Delta_{so}$ and the crystalline splitting $\Delta_c$. The four parameters ($p_\parallel,p_\perp,\Delta_{so},\Delta_c$) can be determined via the four measured values of the $g$-factors, Eq.~\eqref{g:exper:anis}, and the band gap energy $E_g=2.352$~eV. 

We use a least squares fit and find the parameter set to be
\begin{equation}
\label{values}
P_\perp = \frac{\hbar p_\perp}{m_0} = 6.55~\mbox{eV}\cdot\mbox{\AA}, \quad P_\parallel = \frac{\hbar p_\parallel}{m_0} = 7.92~\mbox{eV}\cdot\mbox{\AA}, \quad \Delta_{so} = 1.29~\mbox{eV}, \quad \Delta_c=-0.11~\mbox{eV},
\end{equation}
which yields $\cos^2{\vartheta}\approx 0.7$, $\Delta_{le}=1.34$~eV, and $\Delta_{he}=1.26$~eV. The obtained values of the $g$-factors coincide with the experimental data, see Eq.~\eqref{g:exper:anis}, within the numerical accuracy.

In this work, we abstain from a precise fitting of the parameters for all perovskites, since, for their unambiguous determination, DFT+ETB calculations are needed for low-symmetry phases on the theory side, and a detailed analysis of the $g$-factor anisotropy across different perovskites on the experimental side.

\FloatBarrier

\section{Additional experimental data for MAP\lowercase{b}I\texorpdfstring{$_3$}{3}}

\begin{figure}[h]
	\begin{subfigure}{0.398\columnwidth}
	\centering
		\includegraphics[width=1\columnwidth]{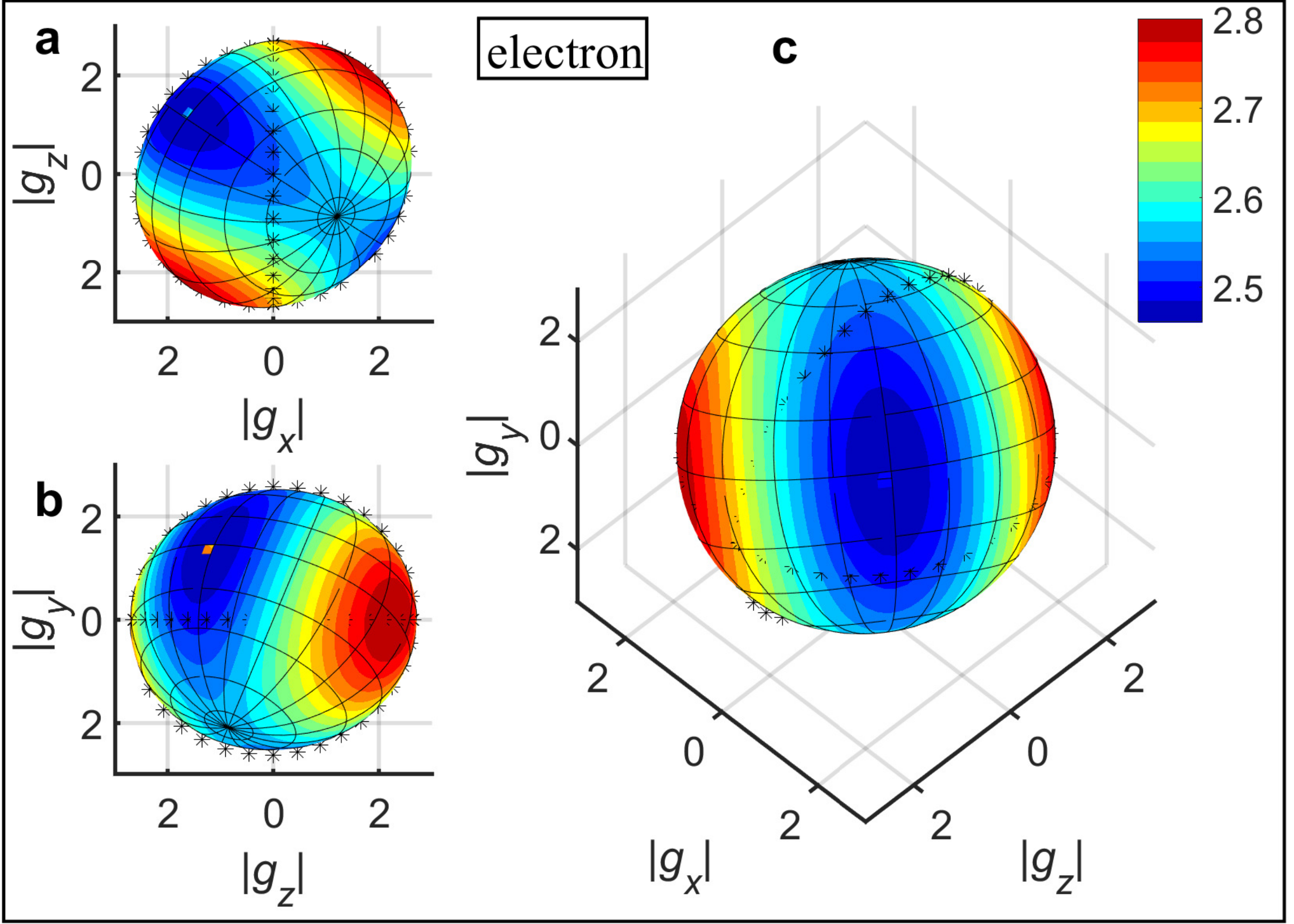}
	\end{subfigure}
	\quad
	\begin{subfigure}{0.38\columnwidth}
		\centering
		\includegraphics[width=1\columnwidth]{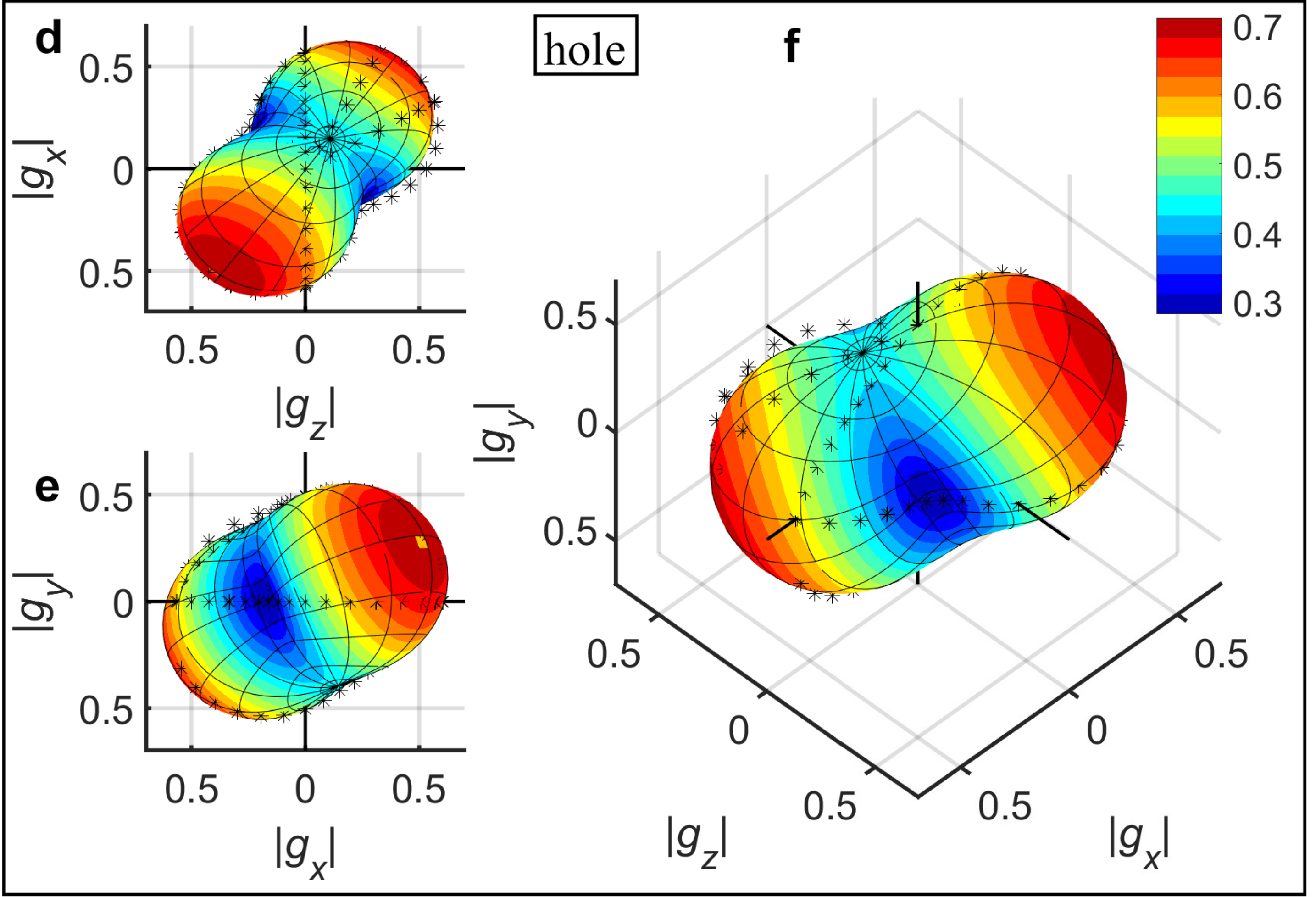}
	\end{subfigure}	
	\caption{\textbf{$g$-factor tensors of the MAPbI$_3$ crystal measured by TRKR at $T=7$~K}. \textbf{a-c}, Electron $g$-factor tensor.  \textbf{d-f}, Hole $g$-factor tensor. Stars are experimentally measured data points.}
	\label{fig:MAPbI_Spheres}
\end{figure}

Three dimensional presentations of the electron and hole $g$-factor tensors measured by TRKR on the MAPbI$_\mathbf{3}$ crystal are given in Fig.~\ref{fig:MAPbI_Spheres}. 

\begin{figure}[h]
\centering
\includegraphics[width=0.85\textwidth]{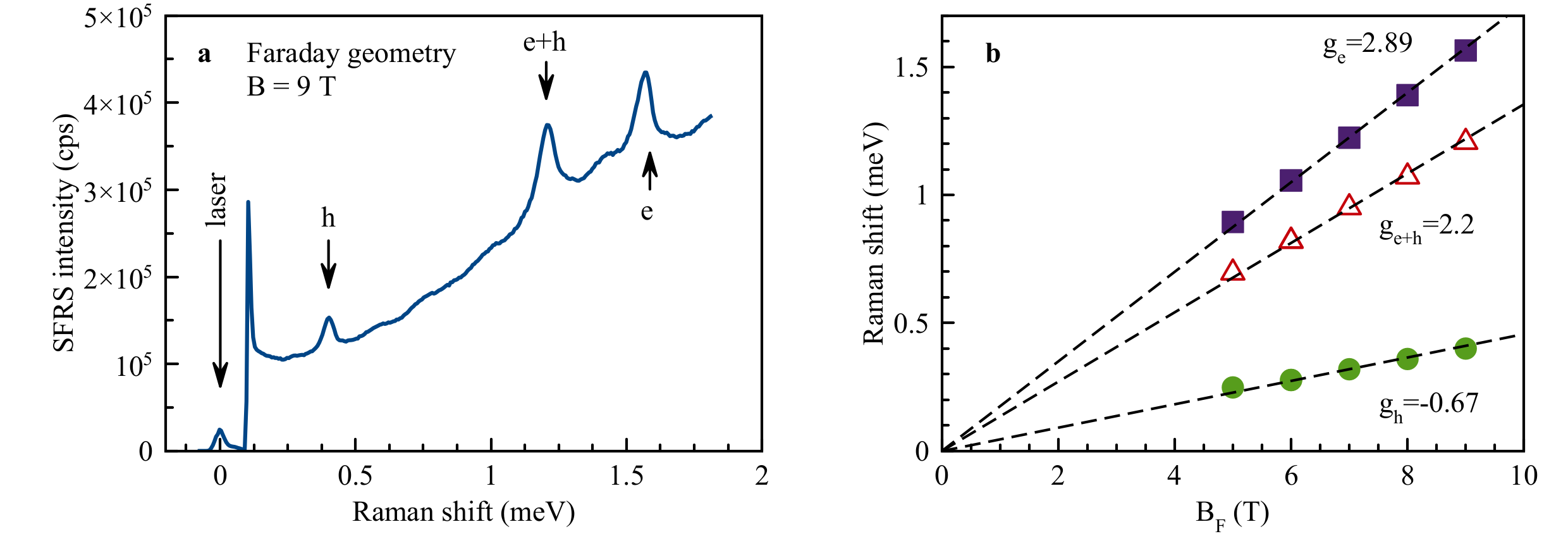}
\caption{\textbf{Spin-flip Raman scattering on the MAPbI$_3$ crystal.} 
\textbf{a}, SFRS spectrum measured in Faraday geometry at $B=9$~T and $T=1.6$~K. Excitation/detection polarization is $\sigma^+$/$\sigma^-$.
\textbf{b}, Magnetic field dependence of the Raman shifts of the hole (green), electron (purple) and combined e+h (red) symbols. The experimental data are shown by symbols, the lines are linear fits.} 
\label{fig:sfrs_MAPI}
\end{figure}

The SFRS measured on the MAPbI$_3$ crystal is shown in Fig.~\ref{fig:sfrs_MAPI}a. Similar to the FA$_{0.9}$Cs$_{0.1}$PbBr$_{0.2}$I$_{2.8}$ crystal in Fig.~\ref{fig:FAPI}a, three Raman lines can be well resolved. They correspond to the hole (h), electron (e) and combined (e+h) spin-flip processes. The magnetic field dependences of their Raman shifts are given in Fig.~\ref{fig:sfrs_MAPI}b.

\FloatBarrier

\FloatBarrier

\end{document}